\documentclass[prb,twocolumn,floatfix,superscriptaddress,nobibnotes]{revtex4}
\usepackage{graphicx}
\usepackage{braket}
\usepackage[caption=false]{subfig}
\usepackage{amssymb}
\usepackage{placeins}
\usepackage{natbib}
\newcommand{\bea}{\begin{Eqarray}}
	\newcommand{\eea}{\end{Eqarray}}
\newcommand{\beq}{\begin{equation}}
	\newcommand{\eeq}{\end{equation}}
\usepackage{color}

\begin{document}

\title{Signatures of Majorana Bound States in the Diffraction Patterns of Extended Superconductor-Topological Insulator-Superconductor Josephson Junctions}
\author{Guang Yue$^{\triangledown}$}
\email{gyue@illinois.edu}
\author{Can Zhang$^{\triangledown}$}

\author{Erik D. Huemiller}
\author{Jessica H. Montone}
\author{Gilbert R. Arias}
\author{Drew G. Wild}
\author{Jered Y. Zhang}
\author{David R. Hamilton}
\affiliation{Department of Physics and Materials Research Laboratory, University of Illinois Urbana-Champaign, Urbana, Illinois 61801, USA}
\author{Xiaoyu Yuan}
\author{Xiong Yao}
\author{Deepti Jain}
\author{Jisoo Moon}
\author{Maryam Salehi}
\author{Nikesh Koirala}
\author{Seongshik Oh}
\affiliation{Department of Physics and Astronomy, Rutgers, The State University of New Jersey, Piscataway, New Jersey 08854, USA}
\author{Dale J. Van Harlingen}
\email{dvh@illinois.edu}
\affiliation{Department of Physics and Materials Research Laboratory, University of Illinois Urbana-Champaign, Urbana, Illinois 61801, USA}
\date{\today}
\begin{abstract}
    In an extended superconductor-topological insulator-superconductor (S-TI-S) Josephson junction in a magnetic field, localized Majorana bound states (MBS) are predicted to exist at the cores of Josephson vortices where the local phase difference across the junction is an odd-multiple of $\pi$. These states contribute a supercurrent with a $4\pi$-periodic current-phase relation (CPR) that adds to the conventional $2\pi$-periodic sinusoidal CPR. In this work, we present a comprehensive experimental study of the critical current vs. applied magnetic field diffraction patterns of lateral Nb-Bi$_2$Se$_3$-Nb Josephson junctions. We compare our observations to a model of the Josephson dynamics in the S-TI-S junction system to explore what feature of MBS are, or are not, exhibited in these junctions. Consistent with the model, we find several distinct deviations from a Fraunhofer diffraction pattern that is expected for a uniform sin$({\phi})$ CPR. In particular, we observe abrupt changes in the diffraction pattern at applied magnetic fields in which the current-carrying localized MBS are expected to enter the junction, and a lifting of the odd-numbered nodes consistent with a $4\pi$-periodic sin$(\phi/2)$-component in the CPR. We also see that although the even-numbered nodes often remain fully-formed, we sometimes see deviations that are consistent with quasiparticle-induced fluctuations in the parity of the MBS pairs that encodes quantum information.
\end{abstract}

\maketitle
\def\thefootnote{$\triangledown$}\footnotetext{These authors contributed equally to this work}
\section{Introduction}
Topological insulators are a new class of materials that are categorized by their topological order instead of by conventional Landau order parameter symmetry, extending traditional classification standards\cite{Qi2011}. These materials have attracted much research interest in the field of condensed matter physics, not only due to their exotic properties but also because they are  potential candidates for hosting Majorana bound states (MBS) which are expected to obey non-Abelian statistics, leading to possible applications in building fault-tolerant topological quantum computers \cite{Kitaev_2003, Kitaev2001}. In the last decade, a number of schemes for creating MBS using topological insulators have been proposed.  Most relevant to this work, Fu and Kane proposed \cite{Fu_2008} a geometry in which two conventional s-wave superconducting electrodes are put into contact with the surface of a 3D-topological insulator, forming a lateral superconductor-topological insulator-superconductor (S-TI-S) Josephson junction. They predict the nucleation of a pair of counter-propagating Majorana modes in the proximitized gap between the electrodes when the the phase difference is an odd-multiple of $\pi$. Due to the topological protection of the surface state on the TI that prevents back-scattering, these states can fuse and carry a supercurrent across the junction through tunneling of quasiparticle single electrons \cite{Beenakker_2013}.  This current exhibits a current-phase relation (CPR) with $4\pi$-periodic sin$(\phi/2)$, which is distinctly different from the usual sin$(\phi)$ Josephson junctions in which supercurrent is solely carried by Cooper pairs with charge unit of $2e$. This paper also showed that a single localized MBS could be nucleated at the center of a trijunction of superconducting islands on a TI in which the relative phases of the islands are controlled to effectively create a phase vortex.  In this case, the partner Majorana state is delocalized in the proximity region around the electrodes.  We previously considered a theoretical model of a 2D system of $p_x+ip_y$-superconductors in which such delocalized Majorana states also occur and mapped out their spatial distribution\cite{Abboud}.

In an extension of this proposal, Potter and Fu \cite{Potter_2013} considered a similar lateral junction geometry but with two primary differences:  (1) a 3D geometry in which the edge of the TI is proximitized so that supercurrents can flow on both the top and bottom surface states, and (2) an applied vertical magnetic field that creates a phase difference across the junction. In this system, the phase difference induced by the applied field delocalizes the extended Majorana states and creates pairs of localized MBS on the top and bottom surfaces of the TI. They propose that these MBS are isolated and can only fuse and generate a supercurrent when they move to the ends of the junction and can interact and hybridize along the edge. 

In this paper, we investigate what we believe is a promising S-TI-S Josephson junction platform for creating and controlling MBS that blends aspects of the Fu-Kane and Potter-Fu configurations.  We have previously discussed the key features of this approach and its advantages for manipulating and interacting the MBS for quantum information processing\cite{Hegde2019}, and presented some of the experimental results\cite{Stehno_2016,Kurter_2015}. Here, we focus on the primary experimental tool we have used, measurements of the critical current diffraction patterns that reveal information about the current-phase relation of the junctions. We present a comprehensive comparison of our CPR model and accumulated experimental results from many junctions and report on the findings.

The S-TI-S junction geometry we propose provides a promising platform for topological quantum computing with potential advantages compared with other popular platforms based on semiconductor nanowires \cite{Mourik12} and chains of magnetic atoms\cite{Nadj-Perge13}. This includes operation without high magnetic fields, intrinsic topological protection from the surface states of the topological insulator, and the ability to move the MBS for braiding and fusion experiments by the application of phases, currents, and voltages to the junction\cite{Hegde2019}.

A number of groups have studied the Josephson effect in S-TI-S junction systems and reported their findings, and progress has been achieved in understanding the proximity effect at the S-TI interface in such systems. Most of these previous works focus on the I-V characteristics and critical current in response to temperature\cite{Veldhorst_2012}, electric field gating \cite{Cho_2013,Sac_p__2011}, the transport properties of the topological insulator\cite{Zhang_2011}, and comparison of different superconductors \cite{Galletti_2014,Zareapour_2012}. Some used different forms of topological insulators, such as thin films or exfoliated flakes from a bulk single crystal, as the Josephson junction’s weak link barrier \cite{Maier_2012,Qu_2012,Yang_2012,Oostinga_2013,Kurter_2014,Stehno_2016}. However, only a few of these studies focused on the CPR of the S-TI-S junctions that we believe encompasses the key physics of this system
\cite{Williams_2012,Sochnikov_2013,Kurter_2015,Sochnikov_2015,Wiedenmann_2016}.  Further, the experimental results presented have only hinted at the existence of a $4\pi$-periodic Josephson effect in the S-TI-S junctions. 

In this article, we first propose a model for describing how the conventional Josephson junction properties would be modified by the existence of MBS in the S-TI-S junction. The main assumption is that the MBS add a supercurrent contribution to the usual uniform Cooper pair tunneling across the lateral barrier. The primary predictions of this model are: (1) the lifting of odd-numbered nodes in the critical current vs. magnetic field diffraction patterns, yielding an odd-even dependence, and (2) the onset of increased supercurrent in the S-TI-S junction diffraction pattern in applied magnetic fields at which Josephson vortices and the MBS bound to them are expected to enter the junction. We then describe the geometry and fabrication details for our S-TI-S junctions that use Nb for the superconducting electrodes and Bi$_2$Se$_3$ as the topological insulator barrier, and present data characterizing the exponential dependence of the critical current on the barrier gap size that we use as a guide for fabricating junctions. We will then present the main result of the paper: a compilation of measurements of the Josephson diffraction patterns of S-TI-S junctions and an analysis of to what extent these exhibit the expected features of the model.  We follow this with a discussion of several other phenomena that may be capable of producing and/or modifying the node-lifting in the diffraction patterns, in particular, critical current variations along the lateral barrier and transitions between parity states of the MBS. In the case of critical current disorder, we present extensive modeling which shows that although such variations certainly occur in real devices, they are not capable, by themselves, of explaining our observations. Finally, we look closely at deviations from the odd-even variation of the node-lifting and show how they may be produced by the expected parity fluctuations of MBS from quasiparticle interactions. We conclude by outlining further experiments, some already in progress, that are suggested by our picture and which should in principle be able to further verify or challenge our model.

\section{The S-TI-S Josephson junction as a platform for Majorana Bound states}

\begin{figure}[htb]
    \includegraphics[width=1.0\columnwidth]{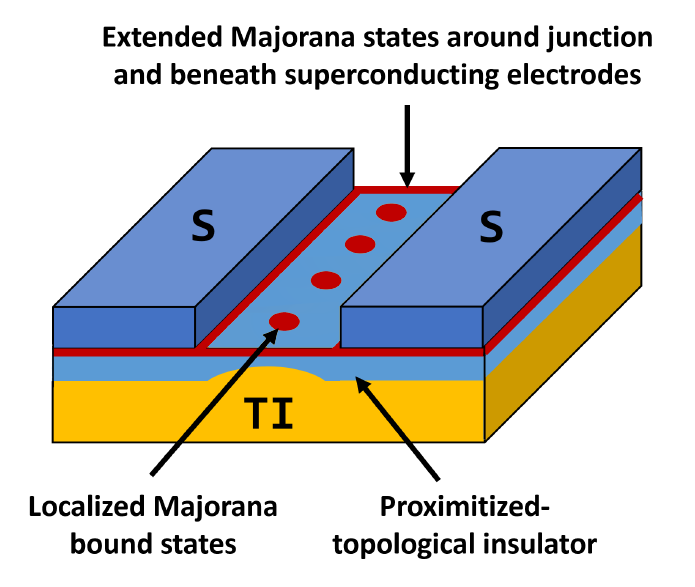}
    \caption {Extended S-TI-S Josephson junction in a weak vertical magnetic field.  Localized MBS exist at locations in the junction where the phase difference is an odd-multiple of $\pi$, with their partners forming delocalized extended MBS in the proximitized region around the junction and underneath the superconducting electrodes.} 
    \label{fig:STIS junction and states}
\end{figure}

The extended (or lateral) S-TI-S Josephson junction device we consider consists of two conventional superconducting electrodes deposited on top of a 3D-topological insulator with a small gap separating them as shown in Fig. \ref{fig:STIS junction and states}. The top surface of the topological insulator beneath and between the electrodes is proximitized by the superconductors, inducing a supercurrent that is confined near the top surface. There is no significant supercurrent on the bottom surface of the TI. The primary path for this supercurrent is the topologically-protected surface state that provides high-transparency conductance with spin-momentum locking that prevents back-scattering, enabling the energy gap to vanish and the subsequent nucleation of MBS. There is also evidence for additional supercurrent contributions from trivial surface states and bulk carriers in some devices. As discussed above, in the absence of an applied magnetic field, and therefore a uniform phase difference across the junction, it is predicted that counter-propagating extended MBS will form in the barrier\cite{Fu_2008,Hegde2019}. Applying a perpendicular magnetic field creates a phase gradient along the width of the junction, nucleating localized MBS at locations in the junction where the phase difference is an odd-multiple of $\pi$. Since MBS have to be created in pairs, the other MBS is delocalized on the surface of the TI in the region around and under the electrodes. This is in contrast to the case of applying the superconductor to the edges of the TI sample, in which a pair of MBS can form on the top and bottom surfaces of the TI at the ends of the Josephson vortex that extends through the thickness of the TI, the case considered by Potter and Fu\cite{Potter_2013}.  

This change in geometry in our devices has several important consequences. First, because of the extended nature of the second MBS in our devices, the localized MBS can entangle with its partner to induce a current at all points along the S-TI-S junction. This is in contrast to the two-layer case in which it has been proposed that the two localized MBS can only contribute a current at the ends of the junction where they can fuse\cite{Potter_2013}. Second, the reduced thickness of the tunneling region reduces the damping of the motion of vortices and enables us to move the Josephson vortices easily and with minimal dissipation. Third, we expect that the protection of the MBS in the junction is enhanced because of its smaller interface with the surrounding materials, but that is yet to be verified by experiment.

In the usual Josephson effect in extended junctions,  the local supercurrent flowing across a Josephson junction depends on the local phase difference across it, and the total integrated critical current of the device modulates with the applied external magnetic field as a result of interference effects.  In a junction with a uniform critical current, uniform applied magnetic field, and a sinusoidal CPR, this gives rise to the characteristic Fraunhofer diffraction pattern\cite{Barone1982} analogous to a single-slit optical interference pattern. Deviations in the diffraction pattern from this functional form can be used to deduce changes in the  CPR.  In particular, in the Fraunhofer pattern, the supercurrent vanishes in sharp nodes at integer values of the magnetic flux threading the junction.  However, the existence of MBS in the junction can create a supercurrent contribution that lifts the odd-nodes, the characteristic signature of a supercurrent with a $4\pi$-periodic CPR. However, it is known that node-lifting effects in diffraction patterns can be also attributed to other factors, including supercurrent density inhomogeneities, vortices trapped in the superconducting films, and flux focusing effects that induce distortion of the applied magnetic field. To address these possibilities, we have supplemented our measurements and modeling of the junction diffraction patterns with additional tests on our junction geometry and properties to ensure that node-lifting effects cannot be mimicked by these other mechanisms.

\section{A model for extended S-TI-S Josephson junctions}
\label{sec:model}
\subsection{The current-phase relation}
In this section, we present a model for the diffraction patterns in S-TI-S lateral Josephson junctions.  Similar to the model discussed by Fu and Kane\cite{Fu_2008} and Potter and Fu\cite{Potter_2013} for a lateral S-TI-S Josephson junction in a weak vertical magnetic field, we assume that localized MBS exist at locations in the junction at which the local phase difference across the junction is an odd-multiple of $\pi$, i.e. at the center of the Josephson vortices in the junction. The number of MBS depends on the applied magnetic field, and in the absence of a current they enter the junctions in pairs from the ends as previously described\cite{Hegde2019}. We further assume that each MBS contributes a supercurrent that superimposes onto the conventional Cooper pair current carried through the TI surface states and bulk states if present. We then integrate the supercurrent across the junction to determine the total supercurrent in the junction as a function of the applied magnetic field, and for each field maximize it with respect to the phase $\phi_0$ at the center of the junction to generate the critical current diffraction pattern.

In our base model, we make several key assumptions that we believe to be the case for our devices: (1) We assume that the phase difference variation along the junction is linear in the applied magnetic field, which requires that both the applied field and the magnetic width of the junction is uniform.  (2) We assume that the supercurrents do not generate any significant magnetic fields in the junction that will distort the applied field. And, (3) we assume that the Josephson CPR is a local property, i.e. that the supercurrent at every location in the junction depends on the local CPR and local phase difference.  This is usually the case in extended junctions, but there can be deviations from this in mesoscopic-scale devices in which the tunneling is not directional, for example, in junctions in which the junction width and length (gap) are comparable\cite{Heida1998}. Deviations from any of these conditions can be calculated but would require corrections to the model. The geometry and parameters of our S-TI-S junctions, discussed in Section IV, are chosen to ensure that these assumptions are met in our experiments.

We therefore propose the following equation to describe the CPR of our S-TI-S junctions 
\begin{equation}
    J(\phi)=J_{c}\left[ \sin (\phi)+ \alpha(\phi)\cdot \sin \left( \frac{\phi}{2}\right)\right]
    \label{Eq:cpr}
\end{equation}
where $J$ is the local supercurrent carried in the junction, $\phi$ is the phase across the junction, $J_{c}$ is the local critical current that scales the magnitude of both the conventional $2\pi$-periodic Josephson supercurrent and the $4\pi$-periodic supercurrent carried by MBS in the junction.  Here, we have assumed that the Cooper pair term contributes a sinusoidal contribution to the CPR.  In fact, because some of the supercurrent is carried by high-transparency surface states, we expect that there will be higher-order harmonics that induce skewness in the $2\pi$-periodic form\cite{Titov2006}.  These are expected and are a necessary precursor to the observation of zero-energy Majorana states.  We use the sinusoidal form here because our simulations show that the skewness has virtually no effect on the diffraction patterns, although it can and has been seen in direct measurements of the CPR, e.g. in graphene\cite{English2016, Nanda_2017} and in S-TI-S junctions\cite{Kayyalha2020, Surendran2023}. 

The factor $\alpha(\phi)$ characterizes details of the MBS localized on the Josephson vortices, which includes their location, extent, and contribution to the supercurrent.  The exact functional form of $\alpha(\phi)$ will depend on the details of the junction. In this paper, we assume a series of Gaussian peaks with the form
\begin{equation}
\alpha(\phi)=\epsilon \sum_n p_n \left(\frac{1}{\sqrt{2\pi}\sigma}\right) exp\left[{-\frac{(\phi-(2n+1)\pi)^2}{2\sigma^2}}\right]
    \label{Eq:alpha}
\end{equation}
with a phase width $\sigma$ and a magnitude $\epsilon$ that scales the current carried by the MBS relative to that of the Cooper pair current and $n$ indexes the phase location of the MBS. A critical parameter implicit in $\alpha(\phi)$ is the parity $p_n = \pm1$ of the MBS pair that encodes the quantum information for the MBS located at phase $(2n+1)\pi$. We have also used an exponential form for $\alpha(\phi)$, which could be expected since this quantity is essentially the wavefunction of the MBS, but the qualitative features of the calculations are not sensitive to this choice.

In this section, we first assume uniform parity of the MBS, assuming that the MBS always enter the junction with the same parity and do not fluctuate in parity over time.  We will discuss the effects of parity changes at the end of this section.

\begin{figure}
    \centering \includegraphics[width=1.0\columnwidth]{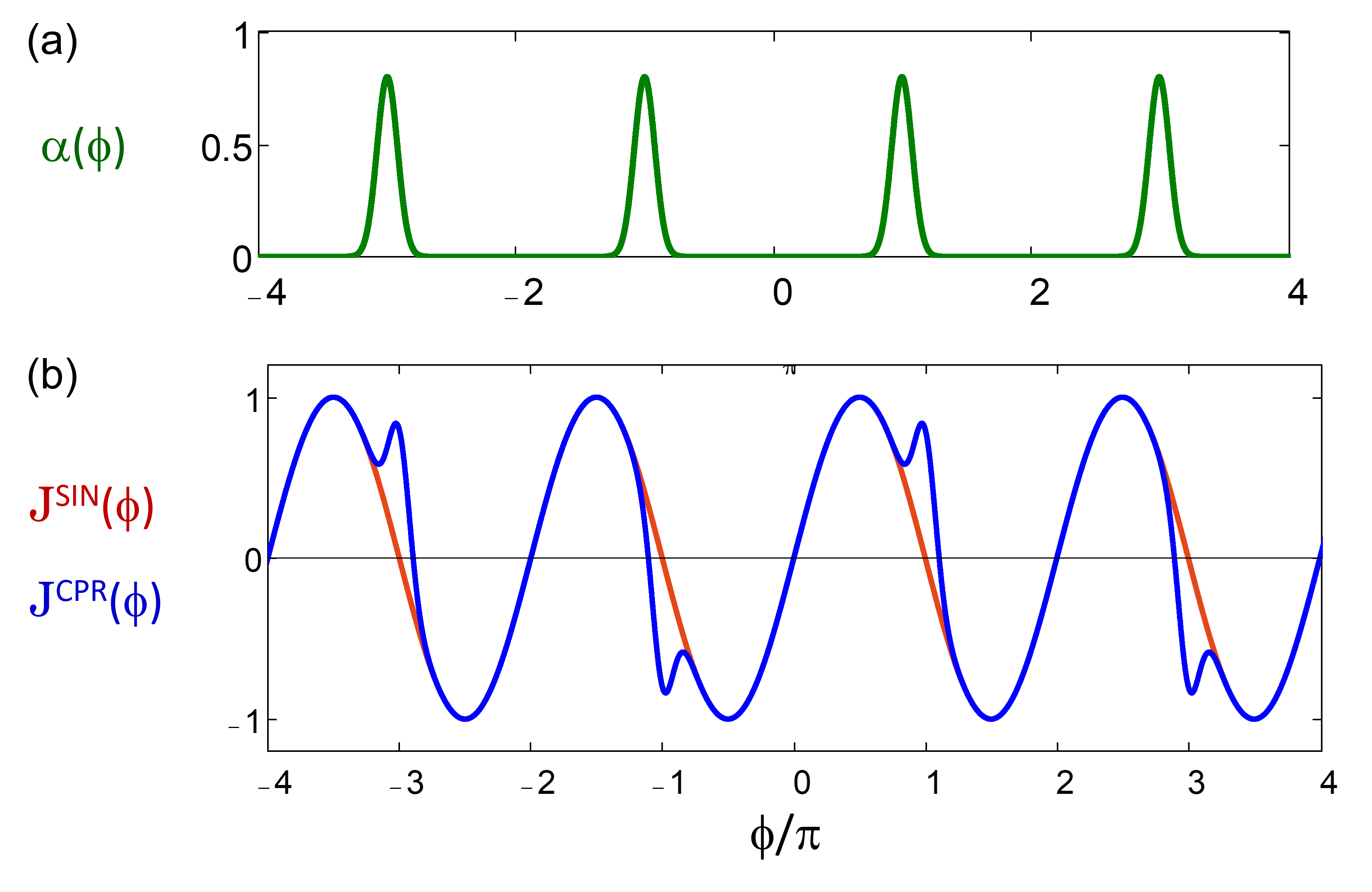}
    \caption{(a) Parameter $\alpha(\phi)$ that characterizes the location and shape of the MBS in an extended S-TI-S junction. (b) The resulting CPR $J(\phi)$ that exhibits localized current spikes with a $\sin(\phi/2)$-dependence from the MBS superimposed on the uniform $\sin(\phi)$ supercurrent, as described in Eqs.\ref{Eq:cpr} and \ref{Eq:alpha}.}
    \label{fig:STIS_cpr}
\end{figure}

A qualitative plot of $\alpha(\phi)$ and the CPR $J(\phi)$ for $\sigma=0.25$ and $\epsilon=0.5$ is shown in Fig. \ref{fig:STIS_cpr}.  The effect of the MBS is the added localized spikes on top of the conventional sinusoidal CPR.  The overall CPR is $4\pi$-periodic.

\subsection{Effect of MBS on the critical current diffraction patterns}
\begin{figure*}[!ht]
    \centering
    \includegraphics[width=2.0\columnwidth]{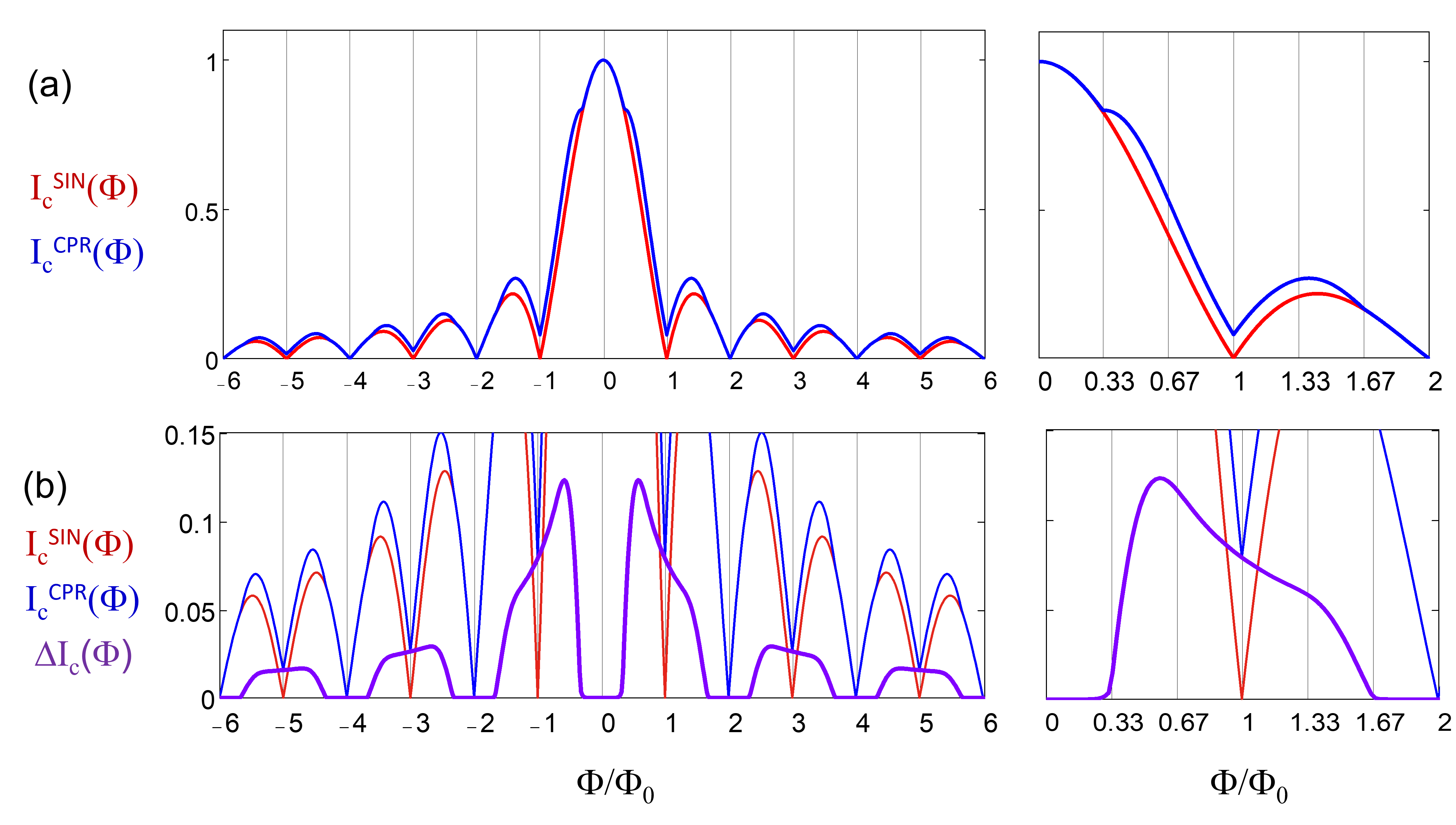}
    \caption{(a) Calculated critical current diffraction pattern $I_c(\Phi)$ vs. magnetic flux $\Phi/\Phi_0$ for an extended S-TI-S Josephson junction. RED: With a uniform sinusoidal CPR that yields the conventional Fraunhofer pattern. BLUE: With the CPR shown in Fig. \ref{fig:STIS_cpr} that includes the contribution from MBS. (b) Change in $I_c(\Phi)$ between the curves in (a) arising from the MBS, plotted in PURPLE. In both of these, we zoom into the region around the first node to highlight the node-lifting contribution to the supercurrent.}
    \label{fig:STIS-diffraction}
\end{figure*}

\begin{figure*}[!ht]
    \centering
    \includegraphics[width=2.0\columnwidth]{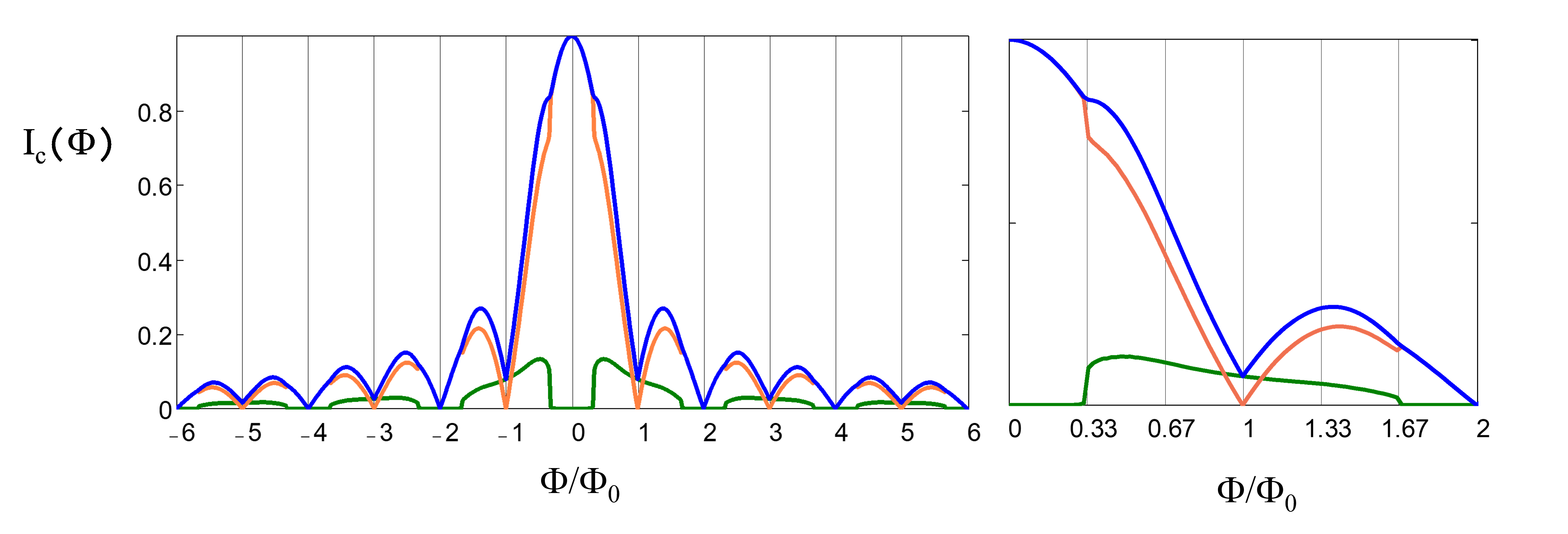}
    \caption{Diffraction patterns for the CPR described in Eq. \ref{Eq:cpr} showing the total current $I_c(\Phi)$ in BLUE, the sinusoidal Cooper pair current in ORANGE, and the $\sin(\phi/2)$ MBS current in GREEN. Abrupt steps occur when the localized MBS enter the junction.}
    \label{fig:STIS-components}
\end{figure*}

Using the Eq. \ref{Eq:cpr} CPR, we can calculate the critical current of the S-TI-S junction in the presence of a magnetic field. When a magnetic field is applied perpendicular to the S-TI-S junction plane, the phase across a uniform junction will change linearly along the direction transverse to the current. Then we can calculate the diffraction pattern: the maximum total supercurrent of the S-TI-S junction as a function of magnetic field. This is plotted in Fig. \ref{fig:STIS-diffraction}(a), where we show the diffraction pattern with MBS compared to the Fraunhofer pattern expected in their absence, both over a wide field range and near the first node in a Fraunhofer pattern.  The critical current in the presence of MBS exhibits a lifting of the odd-numbered nodes in the diffraction pattern and additional structure in the lobes. These features arise as the phase across the junction adjusts to take advantage of the $\sin(\phi/2)$-component. 

As the magnetic field applied to the S-TI-S junction is increased, MBS will enter the S-TI-S junction, producing distinct bump features on the shoulder of the central peak of the diffraction pattern, as shown in the blue curve of Fig. \ref{fig:STIS-diffraction}(a). To highlight these additional features, we plot the difference of the diffraction patterns between the S-TI-S junction with MBS and the conventional uniform Josephson junction without MBS, shown in Fig. \ref{fig:STIS-diffraction}(b). One of the advantages of studying the MBS in our S-TI-S junction system is that on the odd nodes of the S-TI-S junction diffraction pattern, the contribution of the critical current is wholly due to the MBS while the contribution due to the conventional $\sin(\phi)$ CPR of the junction is zero. This makes our experiments highly sensitive to the presence of the MBS states, and provides a way to distinguish these from supercurrent contributions from bulk states and junction critical current disorder. In addition to observing the features on the diffraction patterns, one can also experimentally measure directly the CPR of S-TI-S junctions and search for the spikes in the CPR characteristic of the MBS that are visible in Fig. \ref{fig:STIS_cpr}. 

It is important to note that the incorporation of MBS into the S-TI-S Josephson junctions not only adds an additional supercurrent contribution, but it also modifies the conventional Cooper pair current in the junction. This occurs because the Josephson junction is a coherent quantum device in which phase interference modifies the distribution of supercurrents throughout the junction. The addition of MBS therefore alters the phase profile in the junction, as we will further discuss below, and this changes the relative contribution of the Cooper pairs that exhibit a $2\pi$-periodic sin$(\phi)$-component to the CPR as it adds the $4\pi$-periodic sin$(\phi/2)$-component.  This is demonstrated in Fig. \ref{fig:STIS-components} in which we compare these two contributions. It is interesting that there are abrupt jumps on both the Cooper pair and the MBS supercurrent that arise because the phase profile can adjust to the entry of the localized MBS into the junction to optimize the supercurrent, which minimizes the Josephson coupling energy of the system. 

\begin{figure*}[t]
    \centering
    \includegraphics[width=2.0\columnwidth]{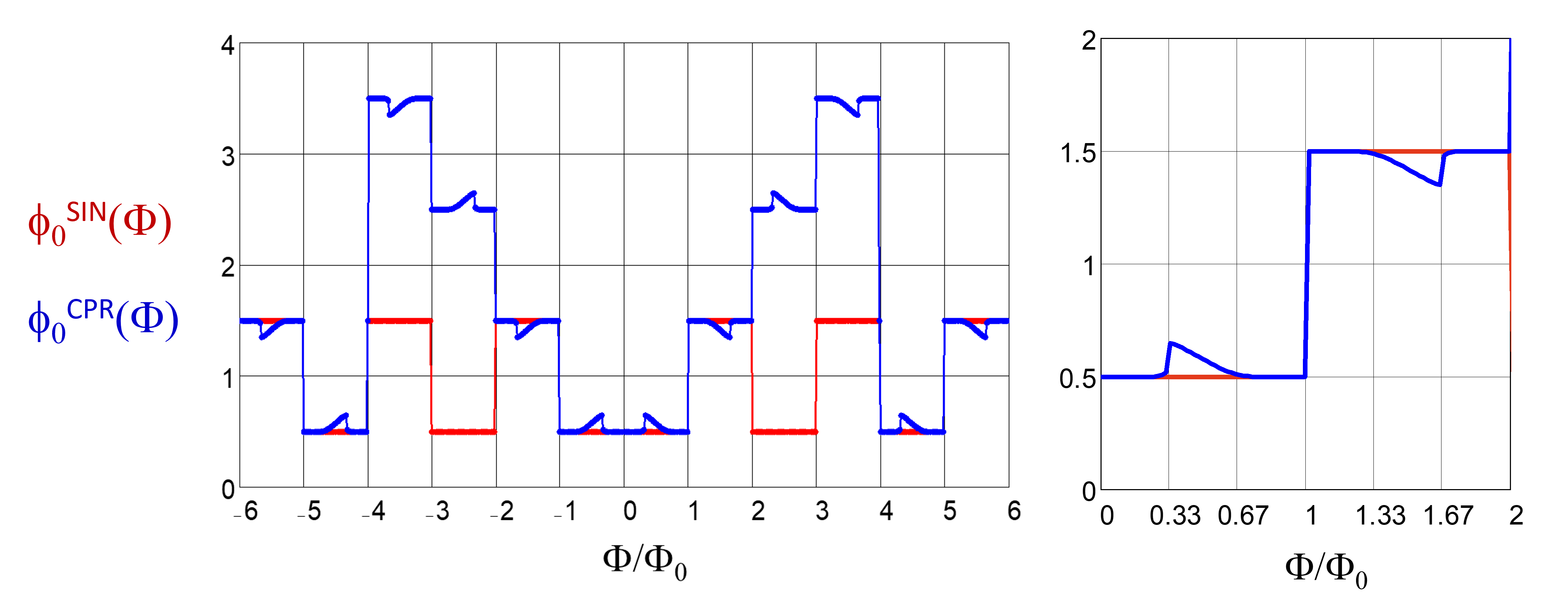}
    \caption{Variation of the phase at the center of the junction vs. magnetic flux $\Phi/\Phi_0$.  RED: With a uniform sinusoidal CPR. BLUE: With the CPR shown in Fig. \ref{fig:STIS_cpr}. Deviations arise at magnetic fluxes for which Josephson vortices and the MBS bound to them enter or leave the junction.}
    \label{fig:STIS-phase}
\end{figure*}

\begin{figure*}[!]
    \centering
    \includegraphics[width=2.0\columnwidth]{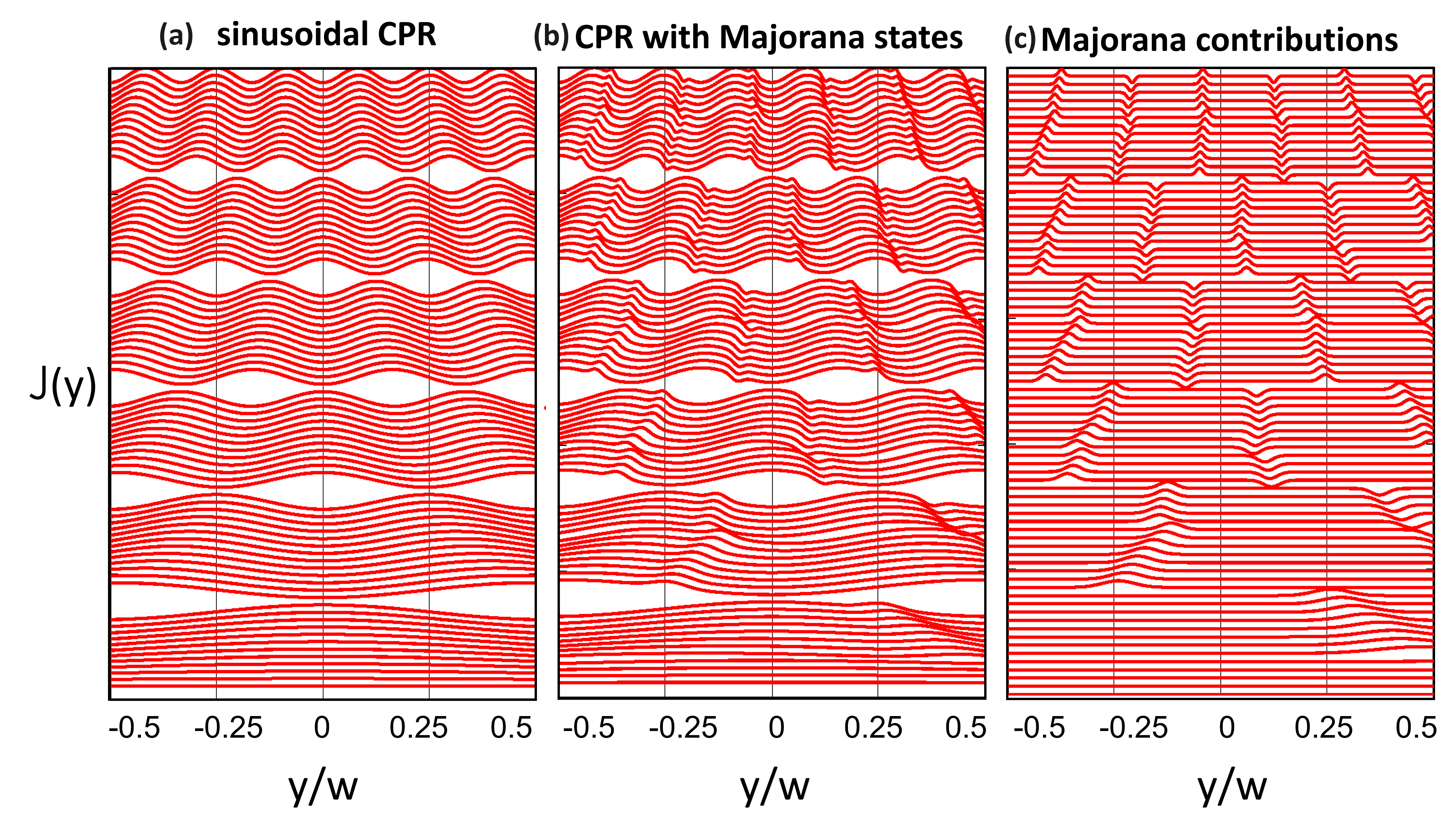}
    \caption{Calculated spatial distribution of the supercurrent across the width of a lateral S-TI-S junction as function of magnetic flux from zero field to 6 $\Phi_0$ in the junction: (a) For a uniform sinusoidal CPR. (b) With a CPR that incorporates MBS in the junction that contribute localized current contributions visible as discrete bumps superimposed  the sinusoidal variation of the current. (c) Showing only the contributions from the localized MBS. The MBS enter the junction from the right, shift to the left when the field changes by one flux quantum, and exhibit small adjustments in position whenever an additional MBS enters the junction.}  
    \label{fig:STIS junction currents}
\end{figure*}

\subsection{Phase variations and the locations of MBS}
In a uniform S-TI-S junction in the short junction limit, with or without MBS, the localized phase $\phi(B,y)$ at magnetic field $B$ will vary linearly along the $y$ direction, which is the direction on the device perpendicular to the current flow. However, the absolute phase is not determined, and needs to be determined by considering the energetics of the Josephson system.  In a single junction, that corresponds to maximizing the total supercurrent that minimizes the Josephson coupling energy. In our calculations, we do this by tracking the local phase difference at the center of the junction, $\phi_0$, from which the phase difference throughout the device and the spatial distribution of supercurrents and location of the MBS can be determined.  

To gain insight into the features in the S-TI-S junction diffraction patterns, we plot $\phi_0$ as a function of magnetic field, shown in Fig. \ref{fig:STIS-phase}. In a uniform Josephson junction with sinusoidal current-phase relations, the phase is always a constant equal to either $\pi$/2 or 3$\pi$/2, depending on the magnetic field, as shown in the RED curve of Fig. \ref{fig:STIS-phase}. The discrete MBS add a 4$\pi$-periodic component to the CPR as shown in the BLUE curve in two ways. First, the phase can vary over 0 to 4$\pi$, and second,  it adjusts to bring the MBS into the junction which enhances the critical current.

The locations of the MBS inside the junction are best visualized by plotting the current in the junction as a function of position, as in Fig. \ref{fig:STIS junction currents}. This is easily calculated using the phase $\phi_0$ determined for each value of applied field. The three panels show (a) the currents for a sinusoidal CPR, (b) the currents for the superposition of the Cooper pair and MBS contributions, and (c) the currents from the MBS contribution alone. It can be seen that the MBS enter the junction from the edge and then move toward the center. It can be seen that they jump in places when the flux crosses an integer number of flux quanta in the junction and the center phase shifts by $\pi$, and also when another MBS enters the junction.  We are currently working to develop a cryogenic Scanning SQUID Microscope (SSM) in a dilution refrigerator to image the junction currents, which is challenging due to the requirement for both high flux sensitivity and high spatial resolution.

\section{S-TI-S junction design and fabrication}

\begin{figure}
    \centering
    \includegraphics[width=1.0\columnwidth]{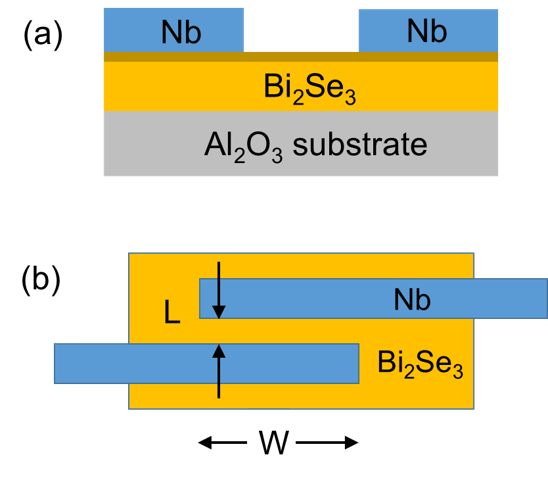}
    \caption{S-TI-S junction geometry:  (a) Side view of lateral junction showing the topological surface state at the S-TI interface that carries the majority of the supercurrent.  (b) Top view of junction identifying the length (tunneling gap) and width of the junction.}
    \label{fig:STIS junction design}
\end{figure}
 
For our S-TI-S junctions, we have adopted a geometry consisting of two narrow parallel superconductor electrodes on top of a topological insulator film, as shown in Fig. \ref{fig:STIS junction design}. Here we define the separation between the niobium electrodes as the length $L$ of the S-TI-S junction, and the extent of the electrodes in contact with the TI film as the width $W$. The critical current of the S-TI-S junctions at zero magnetic field and without gating depends on the length (tunneling gap) and width of the junction, the mobility of the TI surface states, and the conductance of the interface between the superconductor and the TI. For all of our junctions, $W$ and $L$ are chosen so that critical current of the junctions are in the short-junction limit, and so that $W>>L$ to ensure that the supercurrent is highly directional, minimizing mesoscopic effects from supercurrent trajectories at large angles with respect to the normal to the electrode interface. The widths of the superconducting electrodes are chosen to be comparable to the tunneling gap to minimize both flux-trapping in the superconductors and flux-focusing from the Meissner screening of the electrodes\cite{Stan_2004}. In particular, this simple geometry features a well-defined junction width and gap, creating minimal critical current disorder in our junctions. 

For the superconductor, we use magnetron dc-sputtered or electron-beam evaporated Nb with a typical $T_c$ of 8.5 K.  For the topological insulator, we used high quality Bi$_2$Se$_3$ films to ensure that all our transport results are dominated by the topological surface state contribution.  These films were grown with atomic-layer-by-layer Molecular Beam Epitaxy (MBE) on c-plane sapphire substrates at Rutgers University\cite{Koirala_2015}. An In$_2$Se$_3$-Bi$_{1-x}$In$_x$Se$_3$ buffer layer first deposited on top of the sapphire acts as a well lattice-matched virtual substrate for the final growth of Bi$_2$Se$_3$, which yields a highly crystalline defect-free interface at the boundary. These films have high surface state mobilities and small but finite concentration of bulk carriers. It has been previously demonstrated that they also exhibit a trivial surface state that can be depleted by top-gating\cite{Stehno_2016,Kurter_2014}. 

A series of S-TI-S junctions were fabricated by electron-beam lithography. Niobium superconducting leads were deposited through a photoresist lift-off pattern following a few seconds Ar ion-milling cleaning process to ensure a clean interface between the Nb and TI surfaces. Most of the S-TI-S junction films we studied were of order 40 nm in thickness, although thinner and thicker films have also been used and exhibited no qualitative differences in the observed properties. This supports our belief that the supercurrent properties are dominated by the surface states of the TI. The separation $L$ between the two superconducting leads ranges from 100 nm to 400 nm, and the junction width $W$ ranges from 1-5 $\mu$m.  With these parameters, the junction critical currents in zero magnetic field and at low temperature range from 100 nA to 10 $\mu$A.

Devices were mounted to a sample holder wired using an Al wire wedge bonder and measured in a cryogen-free dilution refrigerator with a base temperature of 20 mK. Transport measurements on the S-TI-S junctions are taken using a four-point configuration to avoid picking up contact resistance. Low-pass filters at low temperature are connected between the device and the measurement cables to filter out noise above 1 kHz.

\section{S-TI-S junction measurements}
\subsection{Critical current density}
\begin{figure}
    \raggedright
    \includegraphics [width=0.9\columnwidth] {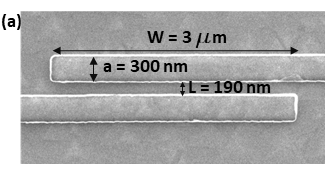}
\par    
   \vspace{0.5cm}
    \includegraphics [width=1.0\columnwidth]{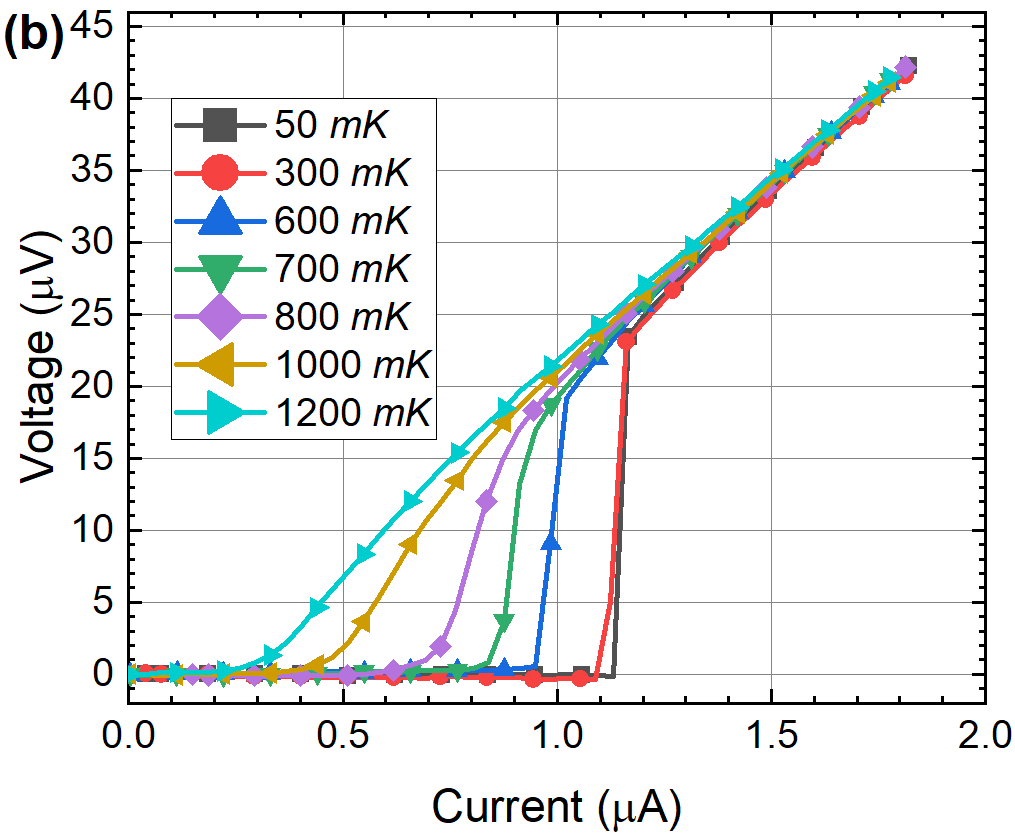}
    \caption{Characteristics of the S-TI-S junctions. (a) Scanning Electron Microscope image of a typical S-TI-S junction with a parallel electrode design style that features a well-defined width, a uniform tunneling gap, and reduced flux-focusing and vortex trapping.  (b) Current vs. voltage characteristic of an S-TI-S junction at different temperatures.}
    \label{fig:IVSEM}
\end{figure}

\begin{figure}
    \centering
    \includegraphics [width=1.0\columnwidth] {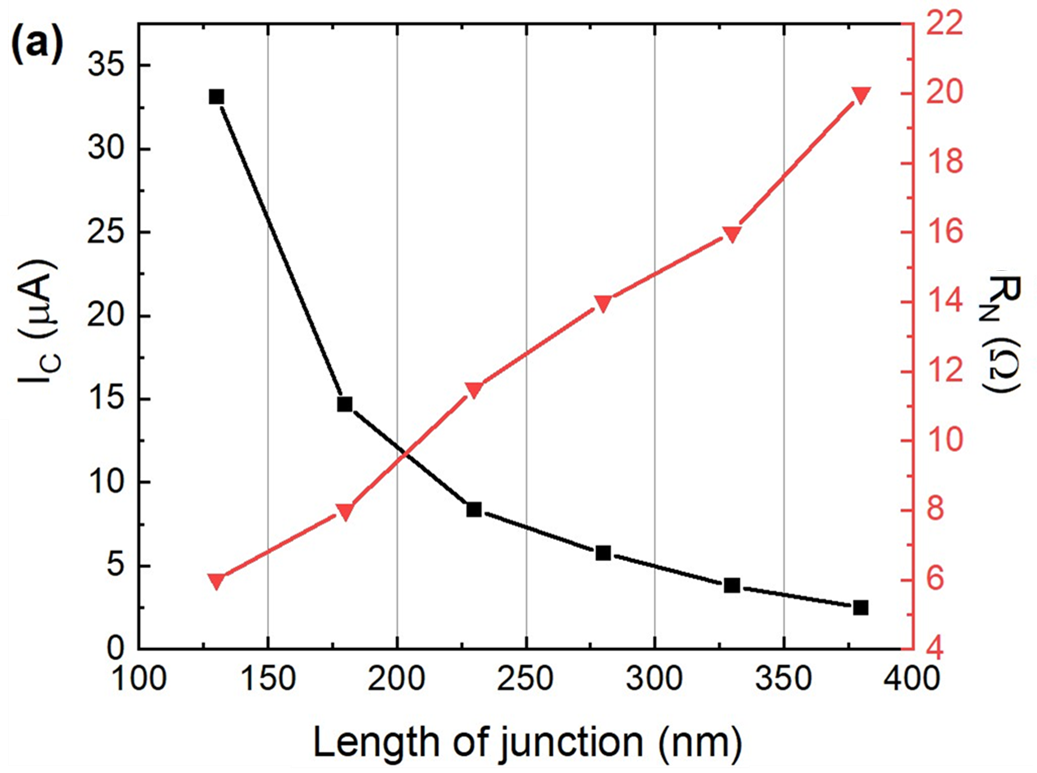}
    \includegraphics [width=0.9\columnwidth] {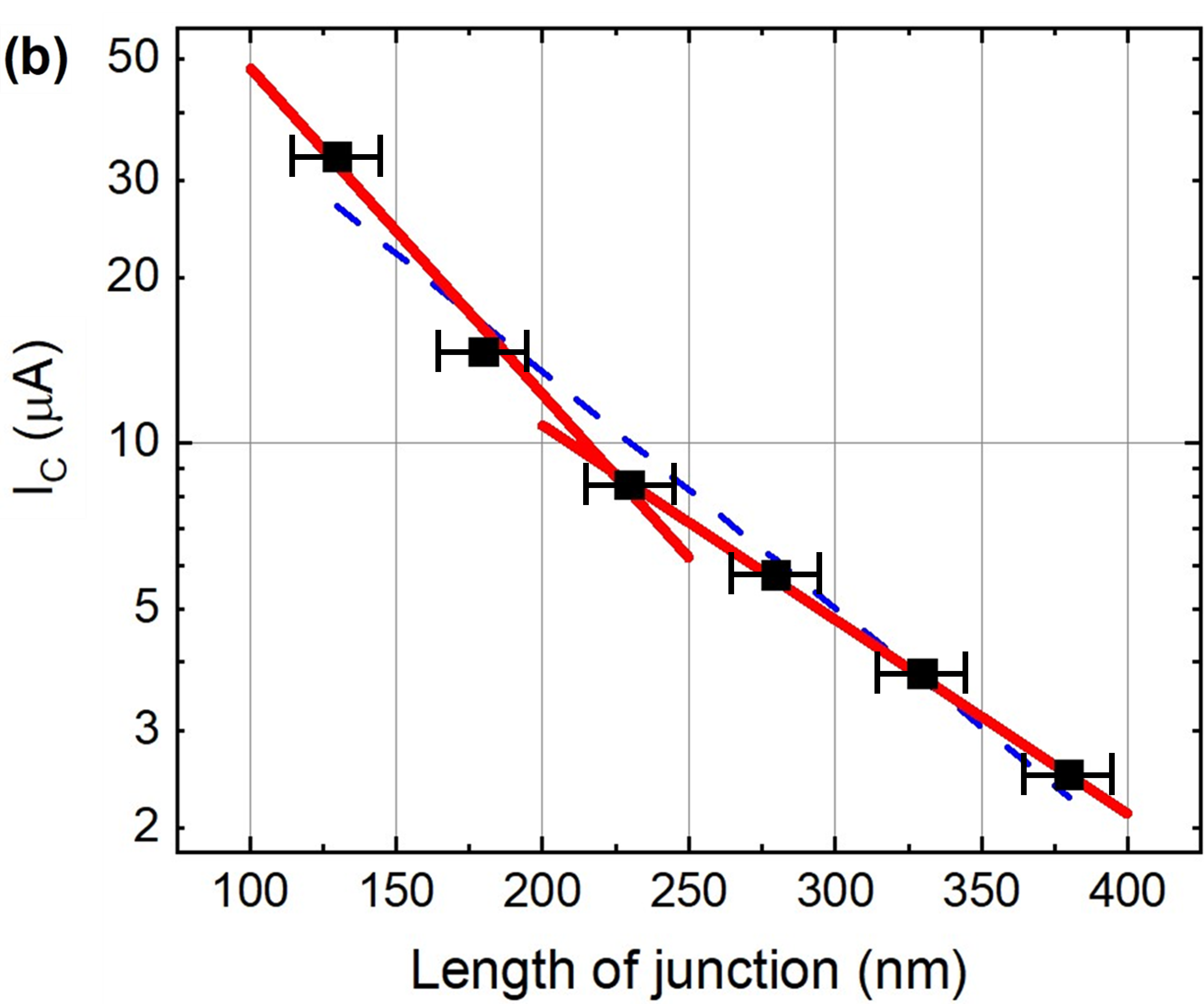}
    \caption{(a) Variation of the critical current and normal state resistance vs. length for a typical S-TI-S junction with W = 3 $\mu m$ (b) $I_c$ vs. length of the S-TI-S junction (BLACK dots) plotted on a log-linear scale. Superimposed are the best fit to a single exponential decay function giving a decay length of 101 nm (BLUE dash), but the data fits best to a double exponential decay function with decay lengths 73 nm and 123 nm (RED solid lines) that is expected because of the two distinct surface states that carry supercurrent. The error bars indicate the error in determining the junction length (gap between electrodes) which we estimate to be $\pm$15 nm.  The error in the measurement of the critical current is small, of the order of the size of the data point.}
    \label{fig:IcRnL}
\end{figure}

A typical S-TI-S junction is shown in Fig. \ref{fig:IVSEM}(a) by Scanning Electron Microscope (SEM) imaging. This junction has a width W of 3 $\mu$m, and gap length L of 190 nm.  The single electrode width is chosen to be 300 nm, comparable to the gap, to minimize flux-focusing and vortex trapping\cite{Stan_2004}. We observe that the length of the S-TI-S junction in the SEM image only has small geometry variation along the width, which indicates that the critical current density along the the width of the junction should be uniform. The current vs. voltage (I-V) characteristic curves of one of our Nb-Bi$_2$Se$_3$-Nb devices are shown in Fig. \ref{fig:IVSEM}(b) at different temperatures. In these I-V curves, the voltage across the S-TI-S junction is zero at small bias current, and when the current is large enough the S-TI-S junction switches into the normal state where a finite voltage appears. The critical current $I_c$ of the S-TI-S junction is defined as the current when it switches from the superconducting state to the normal state. Most of our junctions exhibit a hysteretic I-V curve at zero field and at low temperatures, but this typically disappears as the temperature and/or magnetic field increases and the junction critical current drops. At this time, we do not have a full microscopic model for this behavior -- these junctions have multiple conduction channels that contribute to the supercurrent.  Each arise from different mechanisms and locations in the barrier and exhibit different behavior with respect to temperature, magnetic field, and geometry. The dominant channel is Andreev conduction through the topological surface state that is responsible for both the $2\pi$-periodic current that exhibits skewness from high-transparency states and the $4\pi$-periodic states from localized Majorana states.  There is also a trivial surface state which can be modified by top gating\cite{Stehno_2016,Kurter_2014}, and some bulk conduction that acts like a SNS junction which is probably not affected by gating.  Each of these has a CPR that contributes to the supercurrent and the measured diffraction patterns.  Each also affects the Josephson dynamics and could play a role in limiting the parity lifetime of the Majorana states.  

The junction critical current is strongly sensitive to the distance between the electrodes. The length dependence of one set of our S-TI-S junctions is shown in Fig. \ref{fig:IcRnL}(a) where all the devices are fabricated under the same processes on the same sample chip.  We find that the junction normal state resistance increases linearly as a function of increasing length, extrapolating to zero. Since the critical current of S-TI-S junctions is expected to decay exponentially with the barrier length, we used an exponential decay function, $I(x)=I_0 e^{-\frac{x}{\xi_N}}$, to extract the normal metal coherence length $\xi_N$, which is the characteristic length for the supercurrent decay away from the superconducting electrode that is related to the transparency of the Josephson junction barrier\cite{Tinkahm_2004}. This is best done by plotting the $I_c$ vs. $L$ on a log-linear scale, as in Fig. \ref{fig:IcRnL}(b).  In this set of devices, we actually find the best fit is a superposition of two different values of the normal metal coherence length, with the initial decay being $\xi_N^1=73$ nm and the long-scale decay being $\xi_N^2=123$ nm. This suggests that the supercurrent is carried by two different surface states in Bi$_2$Se$_3$. In addition to the topologically-protected surface state, there is also a trivial surface state due to a 2-dimensional electron gas (2DEG) created by the conduction band bending downward and crossing the chemical potential. This observation is consistent with previous published literature, where the critical current decayed differently as a function of temperature at different gate voltages\cite{Kurter_2014,Oostinga_2013}. We note that the magnitude and decay length of the supercurrent are extremely sensitive to material and fabrication details.  Every set of junctions on a single sample chip that we have characterized shows a similar exponential dependence but the values of the decay parameters vary. 

Another conclusion we can draw from the above critical current vs. length measurement is that the $I_c$ of our junction in the range of 200 nm to 300 nm decays slowly with length, which tells us that a small fluctuation in junction length would only cause a small change in the magnitude of the total supercurrent of the device.  This suggests that critical current disorder is not dominating in this gap-separation regime. To take advantage of this, we fabricate S-TI-S junctions with lengths of about 300 nm for the diffraction pattern measurements in this paper, and expect that this will help minimize the influence of geometry disorder on the results.  We will discuss the effects of such disorder in detail later in the paper.

\subsection{Critical current diffraction patterns} 

Motivated by our initial data and subsequent modeling, the primary results of our study have been to measure and analyze the magnetic field dependence of a large number of S-TI-S junctions. As discussed above, these diffraction patterns should reflect the current-phase relation of the Josephson junctions and provide a test for signatures that might indicate the presence of localized MBS or other phenomena.  It is also well-known that Josephson interferometry is highly sensitive to any deviation in the uniformity of the local junction critical current such as geometric or materials disorder in the junction, spatial variation in the applied magnetic field, and trapped vortices. As we will discuss, it can also provide evidence for the dynamics of parity states in the junctions.  

It is important to point out that we deliberately target junctions with very small critical currents for two fundamental reasons. First, we are seeking resolve the current contributions for localized MBS, so we wish to reduce the integrated contribution from the Cooper pair current.  This is done both by reducing the total critical current, but also by applying a magnetic field that further reduces the critical current by interference effects, while having little effect on the MBS contributions. Secondly, in order use Josephson interferometry to reveal the current-phase relation, we want the extended junction be in the so-called short-junction limit in which the supercurrent depends on the local phase difference in the junction and  the phase difference depends only on the applied magnetic field and is not modified by fields generated by the supercurrents. This criterion is set by the size of the Josephson penetration length $\lambda_J$ relative to the junction width $W$.  In terms of junction parameters, $\lambda_J$ =$(\Phi_0/2\pi\mu_{0}dJ_c)^{1/2}$, where $\mu_0$ is the magnetic susceptibility, $J_c$ is the critical current density, and $d=L+2\lambda$ is the magnetic width of the junction that includes the physical junction gap $L$ and the penetration depth of the superconductor electrode $\lambda$. For our junctions with $L\approx250$ nm and $J_c\approx$ 1 $\mu$A/$\mu$m, $\lambda_J\approx10$ $\mu$m. Compared to our junction width, our junctions are at least marginally in the short-junction limit. 

\subsubsection{Measurement details}

The experimental challenge is to be able to distinguish interesting effects in the current-phase relation from all of the external effects that might mask or mimic them. This challenge is heightened by the magnitude of the critical currents in our junctions which are deliberately designed to be small as discussed above, and the need to look in finite field regimes around the expected nodes to reveal signatures of MBS. Experimentally, the critical current diffraction patterns are extracted from transport measurements of the junction at different magnetic fields. In data analysis, we use two approaches to designate the critical current from the I-V curve at each field: (1) a voltage criterion in which we define the critical current as the current when the voltage across the JJ reaches a certain threshold $V_T$, or (2) a resistance criterion in which we define the critical current as the current when the differential resistance of the junction, as measured by a lock-in measurement of $dV/dI$, reaches a certain threshold $R_T$. Both approaches yield basically the same curve but the resolution depends on the size of the critical current and the noise intrinsic in the measurement circuit.  

\begin{figure}[!b]
    \centering
    \includegraphics[width=1.0\columnwidth]{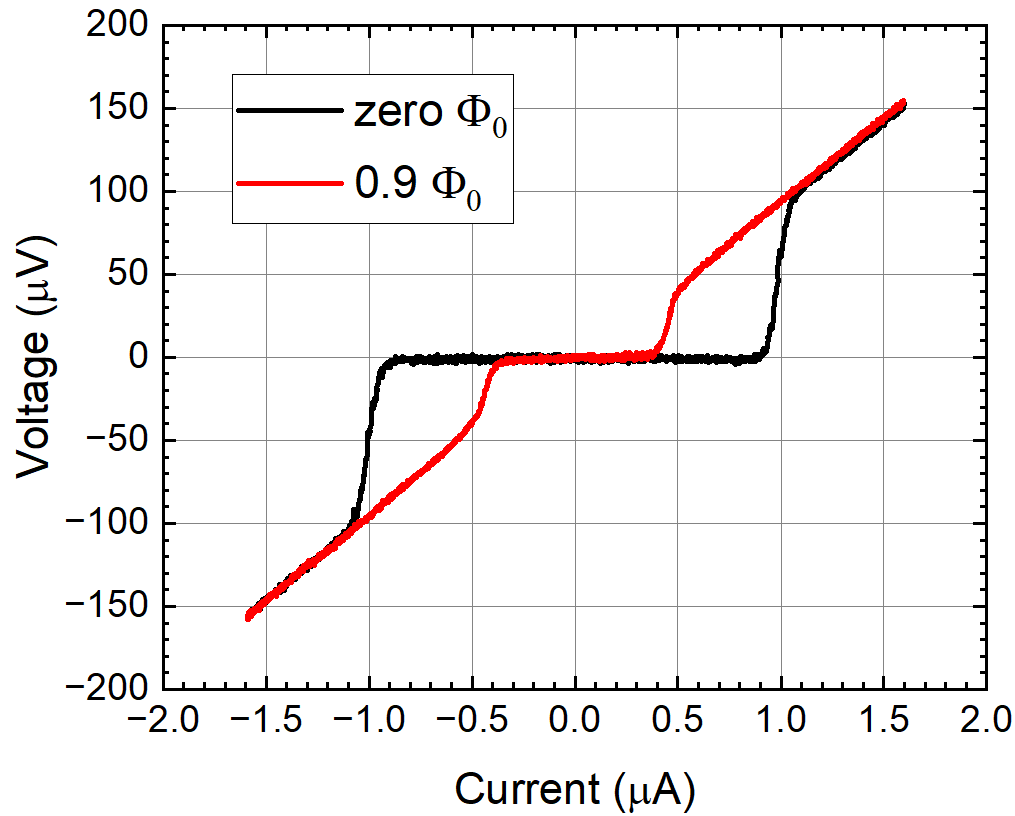}
    \caption{Typical I-V curve for an S-TI-S junction at two values of the applied magnetic field.}
    \label{fig:single_IV}
\end{figure}

A typical I-V curve is shown in Fig. \ref{fig:single_IV} for two different values of the applied flux that corresponds to critical currents of order 1 $\mu$A and 0.5 $\mu$A. The RMS voltage noise level in this data is of order 1 $\mu$V/Hz$^{1/2}$ so we set the threshold voltage just above that, for example at 2 $\mu$V to avoid premature values of $I_c$. It is important to note that this scheme for measuring the critical current requires that the current is sensed at a finite voltage so that the phase is varying in time at the Josephson frequency $f_J=2eV/h$.  For a measurement at $\sim$ 2 $\mu$V as estimated above, that is of order of a frequency of 1 GHz or a period of 1 ns. This plays a critical role in understanding the phase dynamics of the junction and the relevant rate for parity fluctuations to influence the diffraction pattern.  

\subsubsection{Node-lifting}

All of the diffraction patterns in S-TI-S junctions that we have measured, of order 50-100, exhibit interference reminiscent of single-slit optical interference  expected for extended Josephson junctions, with a central peak and multiple-decaying side lobes.  The key distinguishing features are in the field range to which modulations are observed. These features are primarily determined by the uniformity of the junction critical current and the applied magnetic field threading the junction. The structure of the lobes and nodes are also depends on the current-phase relation, as well as local disorder in the junction. In many junctions, we see two distinct deviations from the expected Fraunhofer form sin$(x)/x$.  First, the odd-numbered nodes, particularly the first node, are significantly lifted relative to the other nodes, while the even-numbered nodes remain zero or very small.  Second, on most of the curves there is a distinct glitch in the diffraction pattern curve for magnetic flux close to $\pm$1/3 $\Phi_0$.  Both of these features are consistent with the model for the CPR that we presented above in Sec. \ref{sec:model}.  

\begin{figure}[!hb]
    \centering
    \includegraphics{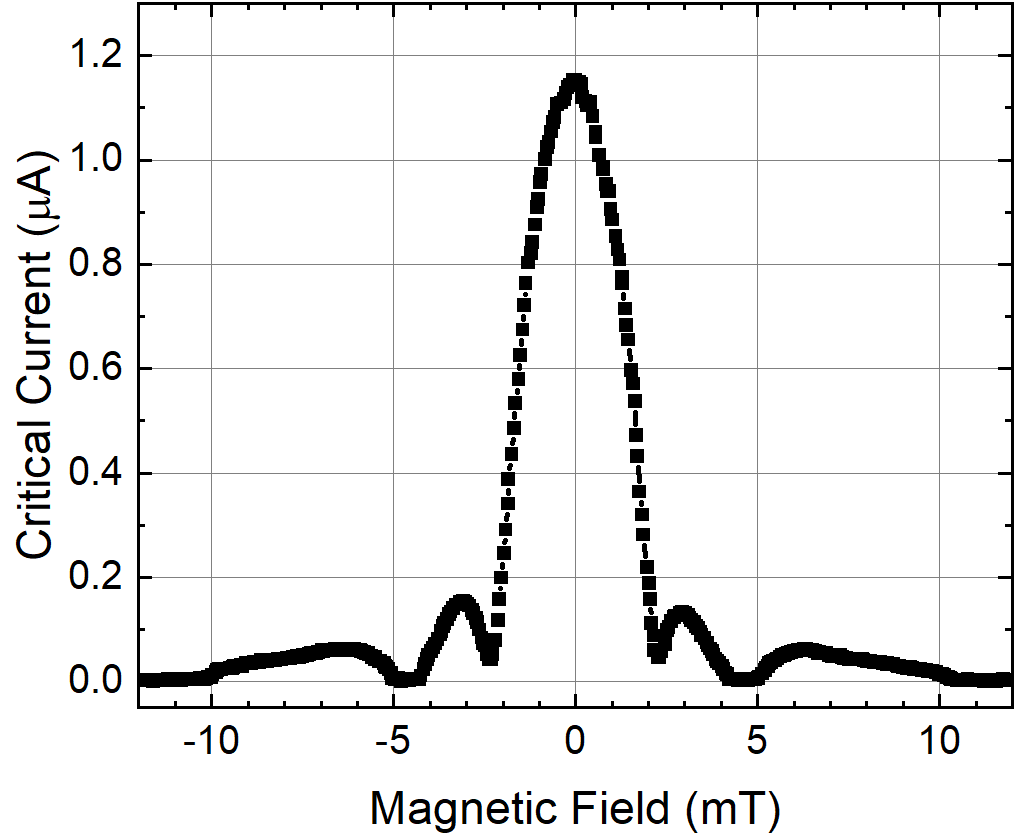}
    \caption{Diffraction pattern showing limited resolution at large fields but clear evidence for a lifted first node and a hard second node. Device dimensions are W = 3 $\mu$m, L = 350 nm.}
    \label{fig:diff_a}
\end{figure}

\begin{figure}
    \centering
    \includegraphics{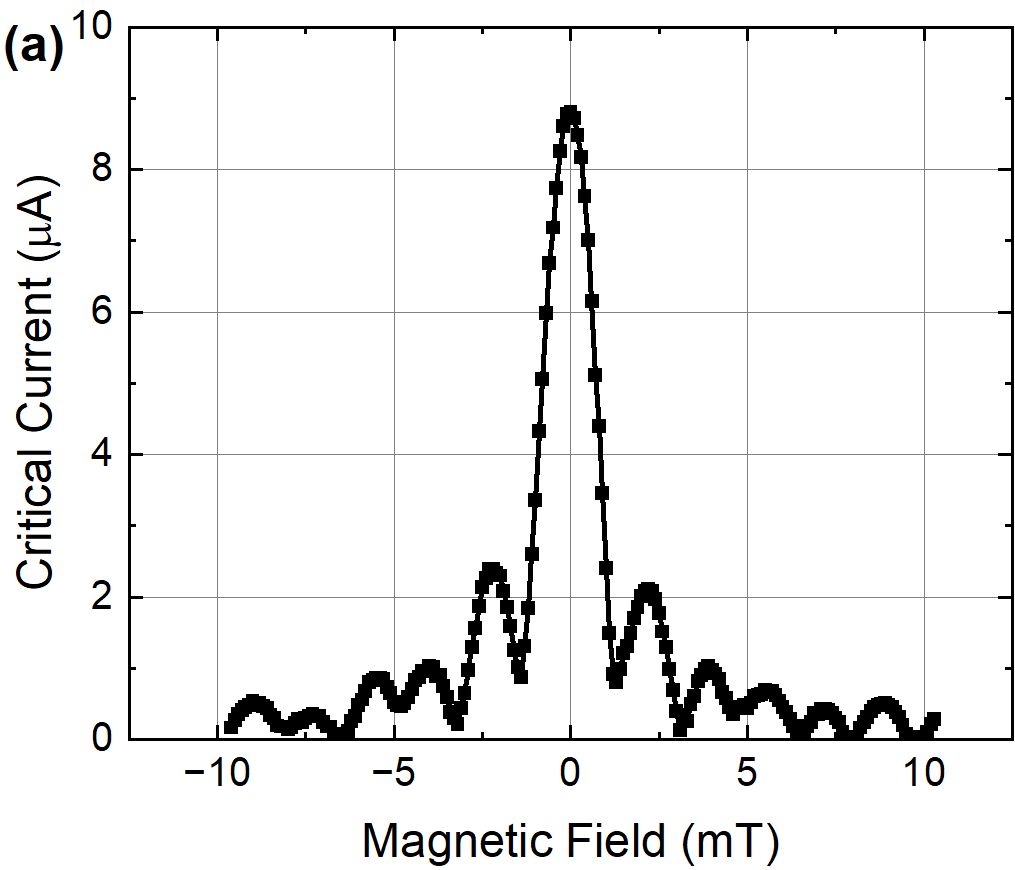}
    \includegraphics{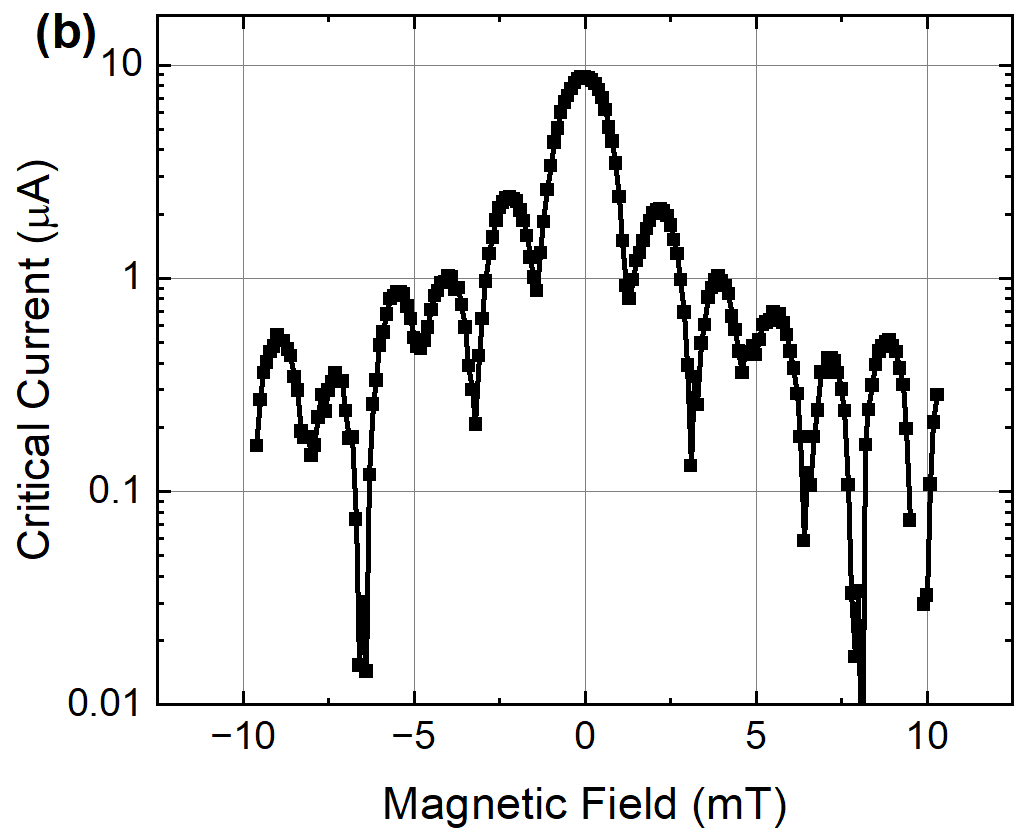}
    \caption{(a) Diffraction pattern showing an odd-even effect with lifted odd-numbered  nodes and hard even-numbered  nodes. (b) Same diffraction pattern plotted on a log scale to enhance the node features. Device dimensions are W = 3 $\mu$m, L = 200 nm.}  
    \label{fig:diff_b}
\end{figure}

To demonstrate these features, we will show a series of diffraction patterns that are representative of the junctions we have measured. All of these were measured at the base temperature of our dilution refrigerator systems that varies between 20 mK and 30 mK on specific cool downs.  In Fig. \ref  {fig:diff_a}, we show a junction that has a limited number of resolvable nodes as the field in increased.  In our experience with Josephson junctions of all types, this usually indicates a non-uniformity in the spatial variation of the local critical current.  Nonetheless, this junction clearly shows that first node is lifted to about $8\%$ of the maximum critical current at zero field of 1.15 $\mu$A, and the second node remains hard at zero critical current. We note that the plateau near the node is simply an artifact of the measurements scheme for extracting the critical current that uses a finite voltage threshold to identify the critical current, resulting in a minimally-detectable critical current.  

In Fig. \ref {fig:diff_b}, we show a junction in which interference features are apparent out to at least 6 $\Phi_0$, suggesting a more uniform junction. In this device we see that the first few odd-nodes are distinctly lifted while the even-nodes remain hard.  This is more clearly seen by plotting the same data on a log scale that highlights the location and depth of the nodes. This odd-even behavior agrees with the prediction of our model and results from interference of the sin$(\phi/2)$-component of the CPR.  Quantitatively, the first three odd-nodes are lifted by approximately $10\%$, $5\%$, and $2\%$.

\begin{figure} 
    \includegraphics[width=1.0\columnwidth]{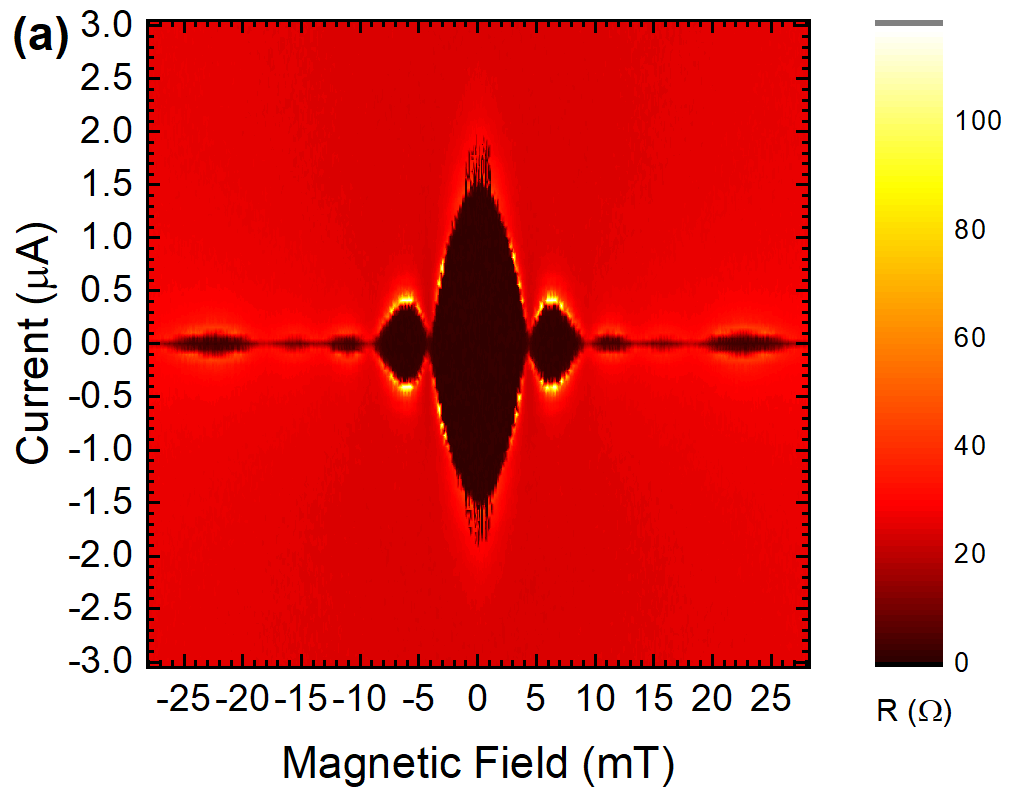}
    \includegraphics[width=1.0\columnwidth]{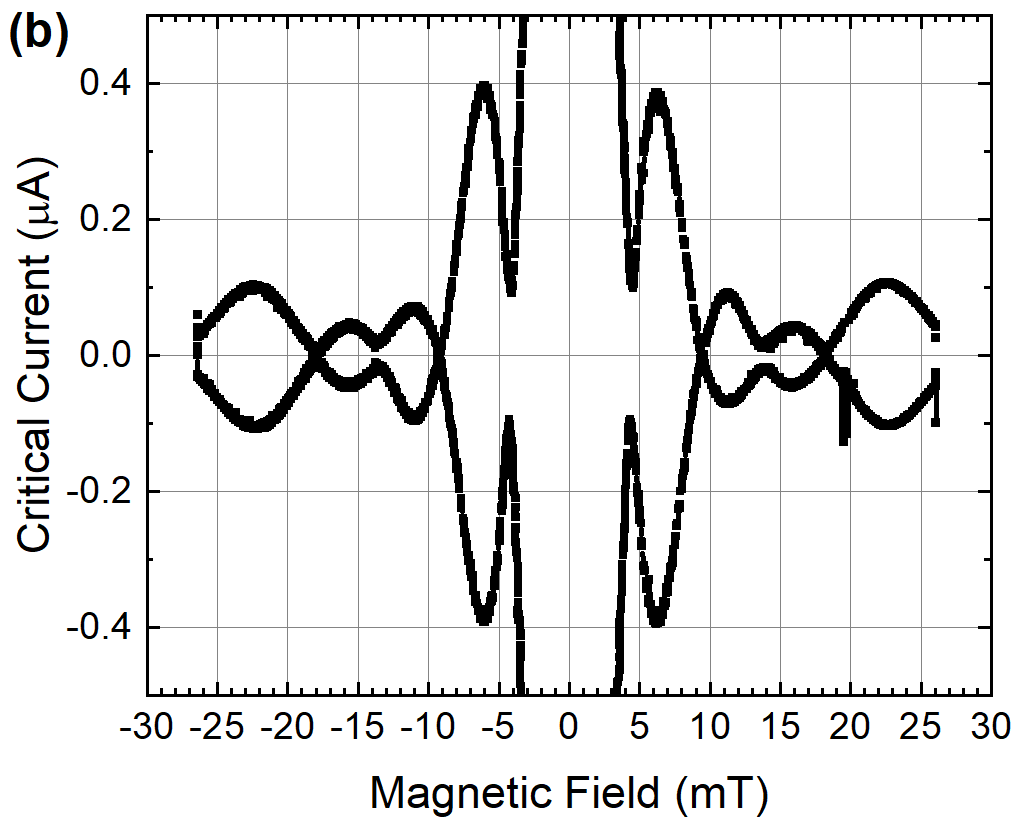}
    \caption{(a) Diffraction pattern plotted in a resistance color plot to highlight the deviations from a Fraunhofer pattern arising from junction disorder. (b) Line plot of the critical current expanded around zero current to highlight  the lifting of odd-numbered  nodes and hardness of even-numbered nodes. Device dimensions are W = 1.5 $\mu$m, L = 100 nm.}
    \label{fig:diff_c}
\end{figure}

Similar odd-even behavior is illustrated in Fig. \ref {fig:diff_c}. In (a), we show the diffraction pattern in a color plot of resistance vs. applied magnetic field, in which black denotes zero resistance which identifies the supercurrent state, and red and yellow indicate finite resistance above the critical current. 
We show this for both polarities of the applied current and the applied field, and the symmetry with respect to these verifies the overall symmetry of the junction critical current.  The deviation from the usual Fraunhofer dependence at higher fields, e.g. the increased lobe at above 20 mT, does show that this junction exhibits some non-uniformity. However, when we zoom-in on panel (b), we see that the node structure again shows lifting of odd nodes and hardness of zero nodes. The first node lifting in this case is around $7\%$ and the third node by around $2\%$.

\begin{figure} 
    \includegraphics{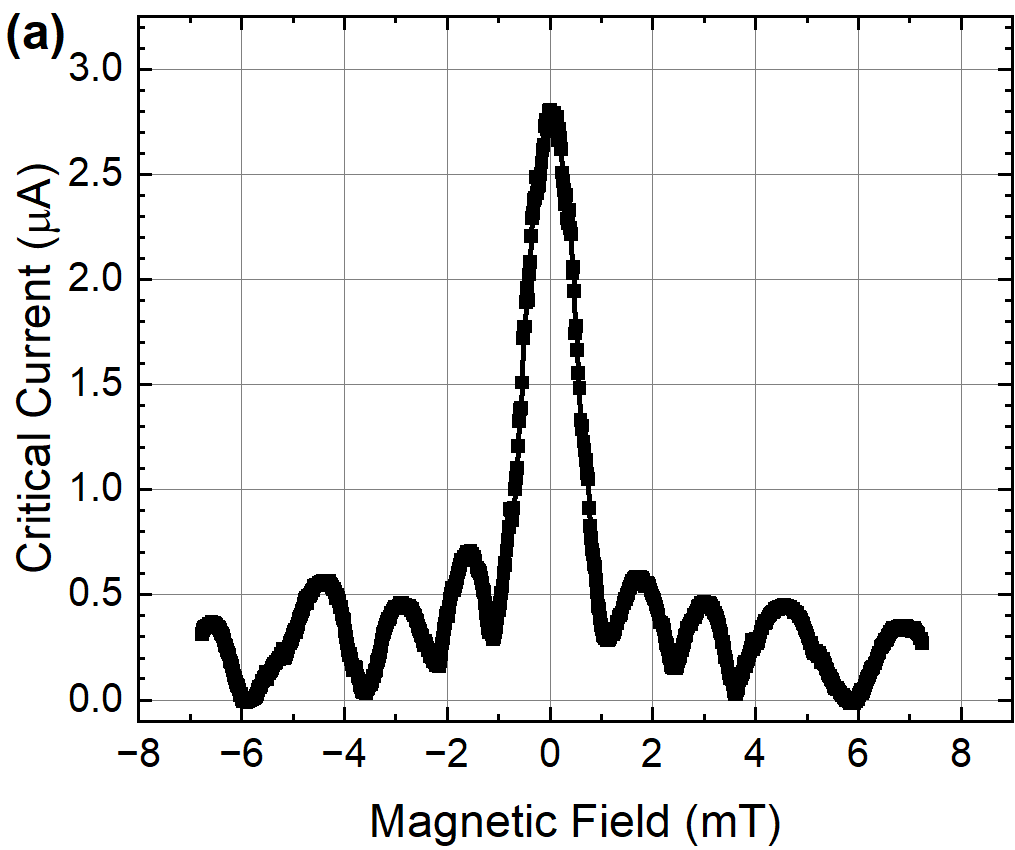}
    \includegraphics{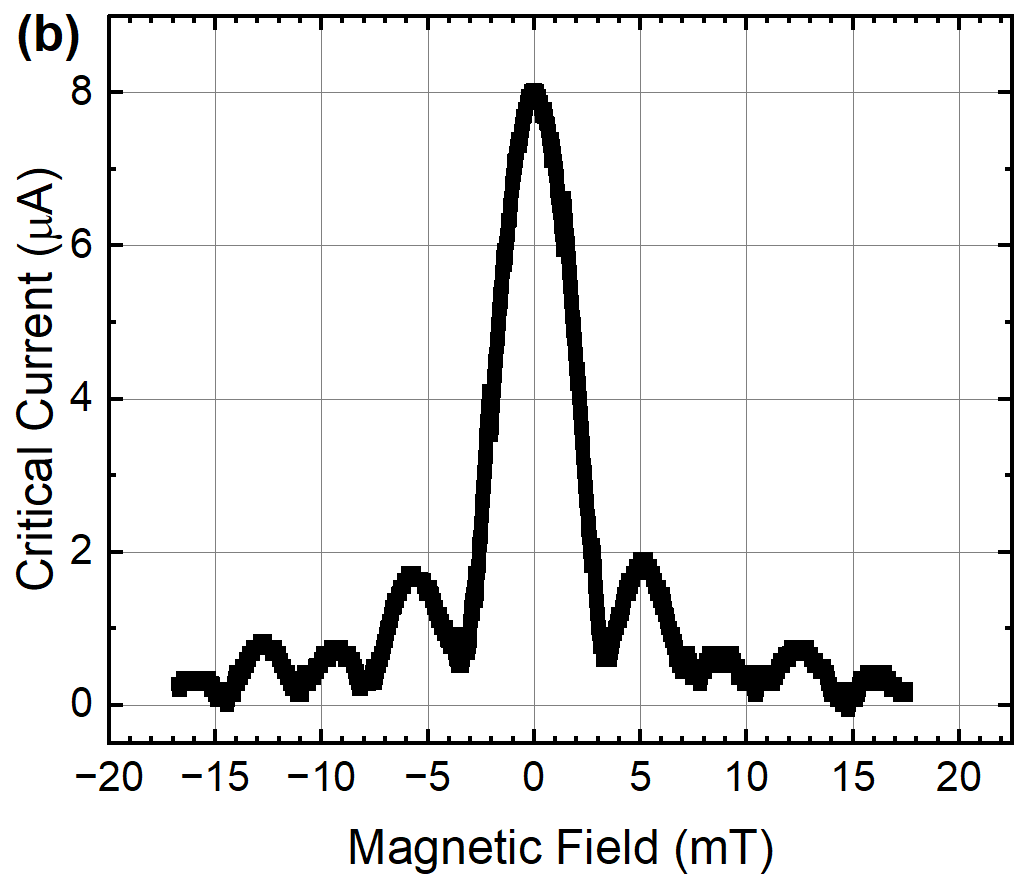}
    \caption{Diffraction patterns showing a variety of node lifting behavior, featuring a substantially lifted first node but also some lifting of all higher order lobes. Device dimensions are (a) W = 3 $\mu$m, L = 290 nm, and (b) W = 3 $\mu$m, L = 200 nm.} 
    \label{fig:diff_d}
\end{figure}

Other devices, for example the two shown in Fig. \ref {fig:diff_d}, show more complicated behavior. Both show a distinct lifting of the first node, again  of order $10\%$, as well as lifting of several other nodes. We particularly note that the second node is not hard, as it is also lifted by a small amount. This is not in agreement with our model, but we discuss how two effects, junction disorder and MBS parity fluctuations, may account for this observation. 

\subsubsection{Entry features}

\begin{figure}
    \centering
    \includegraphics[width=1.0\columnwidth]{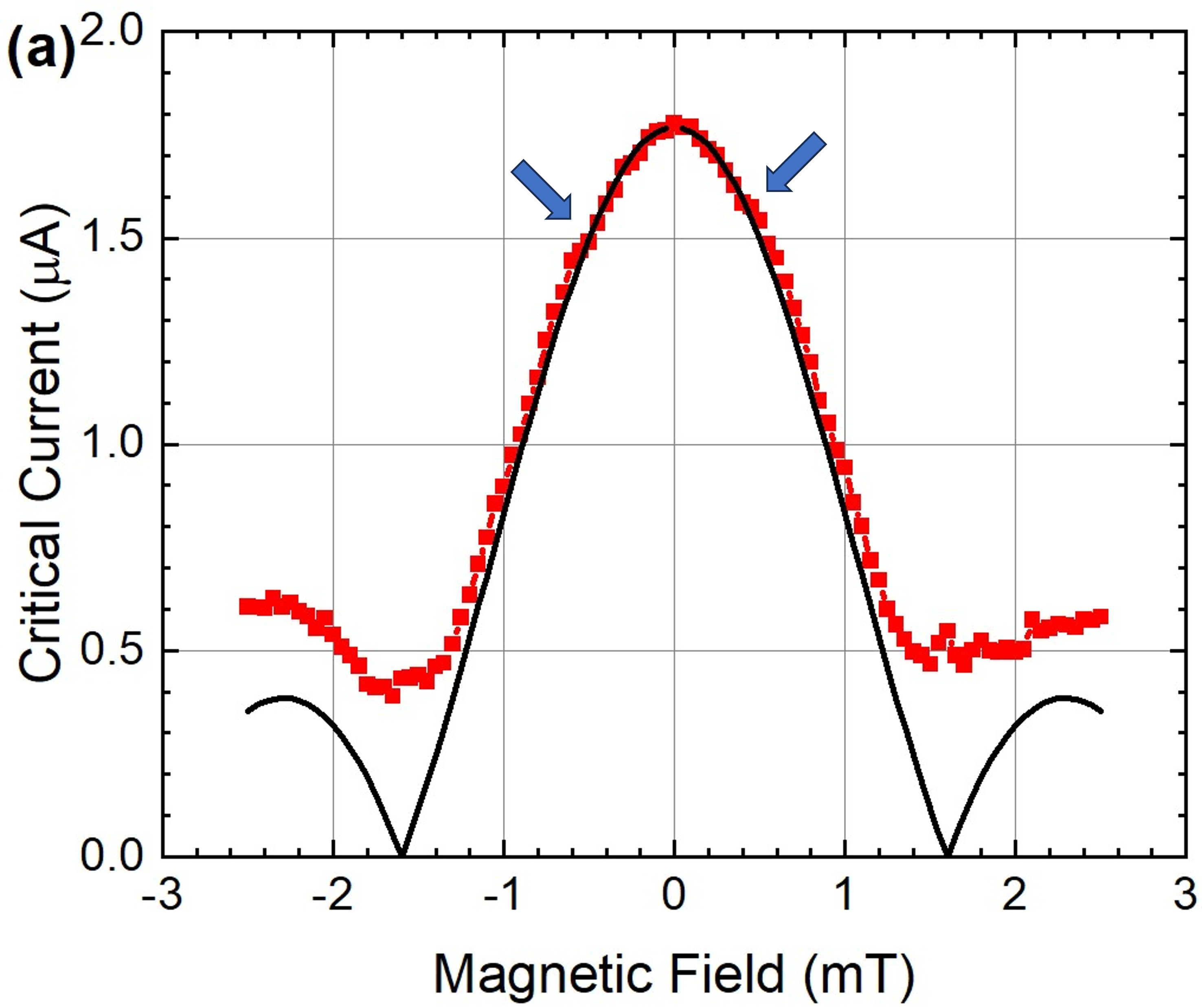}
    \includegraphics[width=1.0\columnwidth]{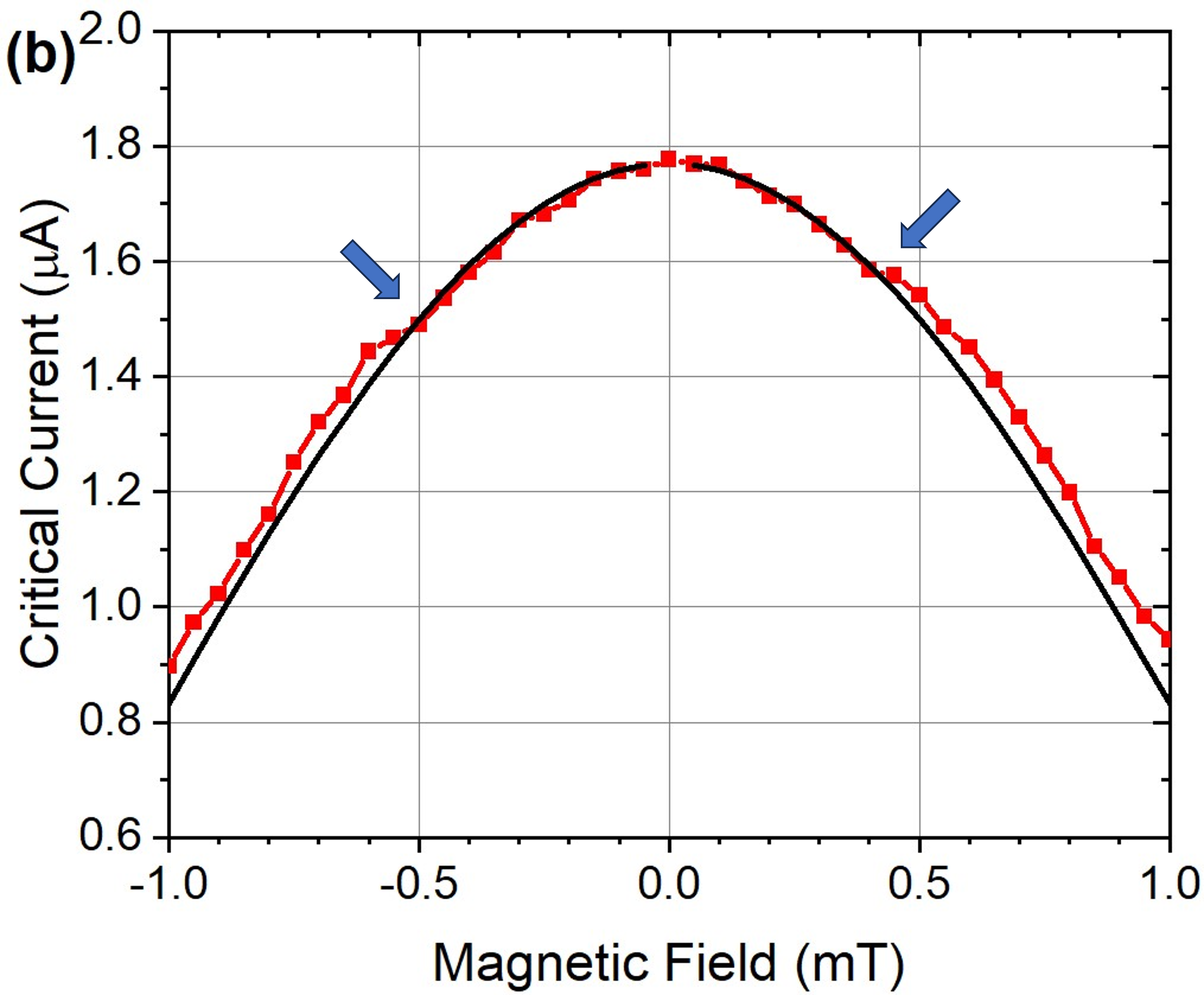}
    \caption{Entering feature on the measured diffraction pattern of an S-TI-S Josephson junction. (a) Diffraction pattern data in RED, simulated diffraction pattern assuming a  $\sin(\phi)$ CPR in BLACK. (b) Zoomed in plot around the shoulder of the diffraction pattern.} 
    \label{fig:entering}
\end{figure}

Other than the odd-even node-lifting effect, we also observe an abrupt vortex entry feature at the shoulder of the diffraction pattern, which we interpret as the entry of the first Josephson vortex and MBS bound to it when the $\phi=\pi$ phase difference appears within the S-TI-S junction.  A close look at the diffraction patterns show a small hint of this feature near the top of the central peak, but the feature is small and requires more careful measurements to resolve.  In Fig. \ref{fig:entering}(a) we can see the diffraction pattern  has node lifting and a small bump on the shoulder of central peak of the curve. The vortex entry feature is further shown in the enlarged plot of Fig. \ref{fig:entering}(b), where the measurement data clearly deviates from the simulation at larger magnetic fields and foreshadows the subsequent node-lifting. This feature is symmetric with respect to the peak of the critical current at zero applied magnetic field. The simulation in Fig. \ref{fig:entering} is a calculated diffraction pattern assuming regular $\sin(\phi)$ CPR.  This feature is small but significant in that it shows an increase in the junction critical current at a distinct field, demonstrating the onset of a localized MBS with excess current. 

\section{Comparison of measurements and model} 
\subsection{Agreements and deviations}
Since what we have presented is only a subset of the junctions that we have measured, it is fair to ask how representative these devices are of the overall measurements we have made.  In that regard, we can summarize our observations as follows:

(1) We have fabricated and measured of order 100 S-TI-S junctions which exhibit measurable diffraction patterns. Some have been measured in detail and analyzed in detail, where as many have only been observed for their qualitative shape. For the purpose of doing statistical analysis on the data, we sample a subset of around tens of devices for detailed analysis shown below.

(2) Of these, most ($>$90$\%$) show a distinct lifting of the first node of height $\sim 5-15\%$ of the maximum zero-field $I_c$. This seems to be a characteristic feature of the S-TI-S junctions that to our knowledge is not normally seen in non-topological Josephson junctions.   

(3) A significant fraction of these, around half, also show a hard or very small second node, consistent with that usually seen in conventional junctions. In the remaining devices, we see a second node that is lifted by a resolvable amount.  We will show below that this is expected due to parity fluctuations of the MBS in the junctions.  

(4) It is challenging to measure the heights of higher-order  nodes due to the low critical currents at those magnetic fields.  However, in the junctions in which we can measure this, we see in a number of cases the lifting of odd nodes and the hardness of even nodes. The junctions we show in Figs. \ref {fig:diff_b} and \ref{fig:diff_c} represent some of the clearest examples of this. 

(5) Some devices show more-or-less random variations of the node heights beyond the first node which is almost always distinctly lifted.  It is known that all Josephson junctions of all types are susceptible to critical current disorder that in general lifts all nodes.  We will analyze the effect of this for our junctions in the next section.

(6) We also observe an abrupt vortex entry feature at the shoulder of the diffraction pattern in many of our junctions.  This often shows up just as a small glitch of the critical current, but careful measurements and data analysis have mapped its shape in detail in some junctions.  

We note that similar node-lifting effects have been seen in previous published literature \cite{Kurter_2015,Williams_2012}, but many of these authors explained this effect with disorder\cite{Cho_2013}, flux focusing effects \cite{Williams_2012} or simply ignored the effect entirely, and none of them showed such consistent odd-even node-lifting effect symmetric around the peak to the fourth nodes.  These observations have been observed in many different forms of Bi$_2$Se$_3$ samples, including MBE-grown thin films involved in this paper and  exfoliated flakes from Bi$_2$Se$_3$ crystals.

\subsection{Effects of critical current density variations}
It is important to address the possibility that non-uniformity in the critical current of Josephson junctions could be  responsible for the node-lifting effects that we see in our experiments. It is rather well-known that such disorder can modify the diffraction patterns of junctions, causing deviations from the Fraunhofer functional form that characterizes the interference effect, in particular causing a smearing out of the sharp nodes at integer values of the magnetic flux threading the junction. In our S-TI-S junctions, this disorder is likely to cause the critical current density $I_c(y)$ of the junction to be non-uniform along the width of the S-TI-S junction (the $y$ direction transverse to the current flow). Our model can also calculate the node-lifting of the diffraction pattern due to the geometry disorder effect. The calculation results show that the geometry disorder in our fabrication can only cause a small and random node-lifting of the diffraction pattern which is quite different from the odd node-lifting feature mentioned above. 

One important aspect of our experiment to test is the dependence of the critical current diffraction patterns on the uniformity of the critical current density across the width of the junction.  Because the local Josephson critical current density is sensitive to the length of the tunneling path and to the contact resistance between the barrier materials and the superconducting electrodes, there can be significant variations in the critical current density.  This can be enhanced by any local defects or inhomogeneities in the conductance of the barrier. It is well-known that these can modify the magnetic field dependence of the critical current of the junction.  In an ordinary Josephson junction with a sinusoidal CPR, the critical current can be simply related to the Fourier transform of the critical current density.  In the case of disorder, the Fraunhofer form of the diffraction pattern for a uniform current density is modified, typically lifting all of the nodes.  Since we are looking at the node-lifting of the diffraction patterns in our S-TI-S junctions as a signature of a $\sin (\phi/2)$ contribution to the critical current that might arise from MBS, it is important to assess whether these effects might be described by non-uniform critical current density effects. 

To that end, we have carried out a careful evaluation of our junctions to estimate the expected critical current variations in our junctions, and also  performed an extensive simulation of the effects of critical current density disorder on the diffraction patterns. Examination of our devices by SEM shows an average variation in the spacing between the electrodes to be no more than 15 nm for a typical barrier width of 300 nm, corresponding to a $5\%$ variation. For devices in this range, the change of the critical current is also about $5\%$. 

\begin{figure}
    \centering
    \includegraphics[width=1.0\columnwidth]{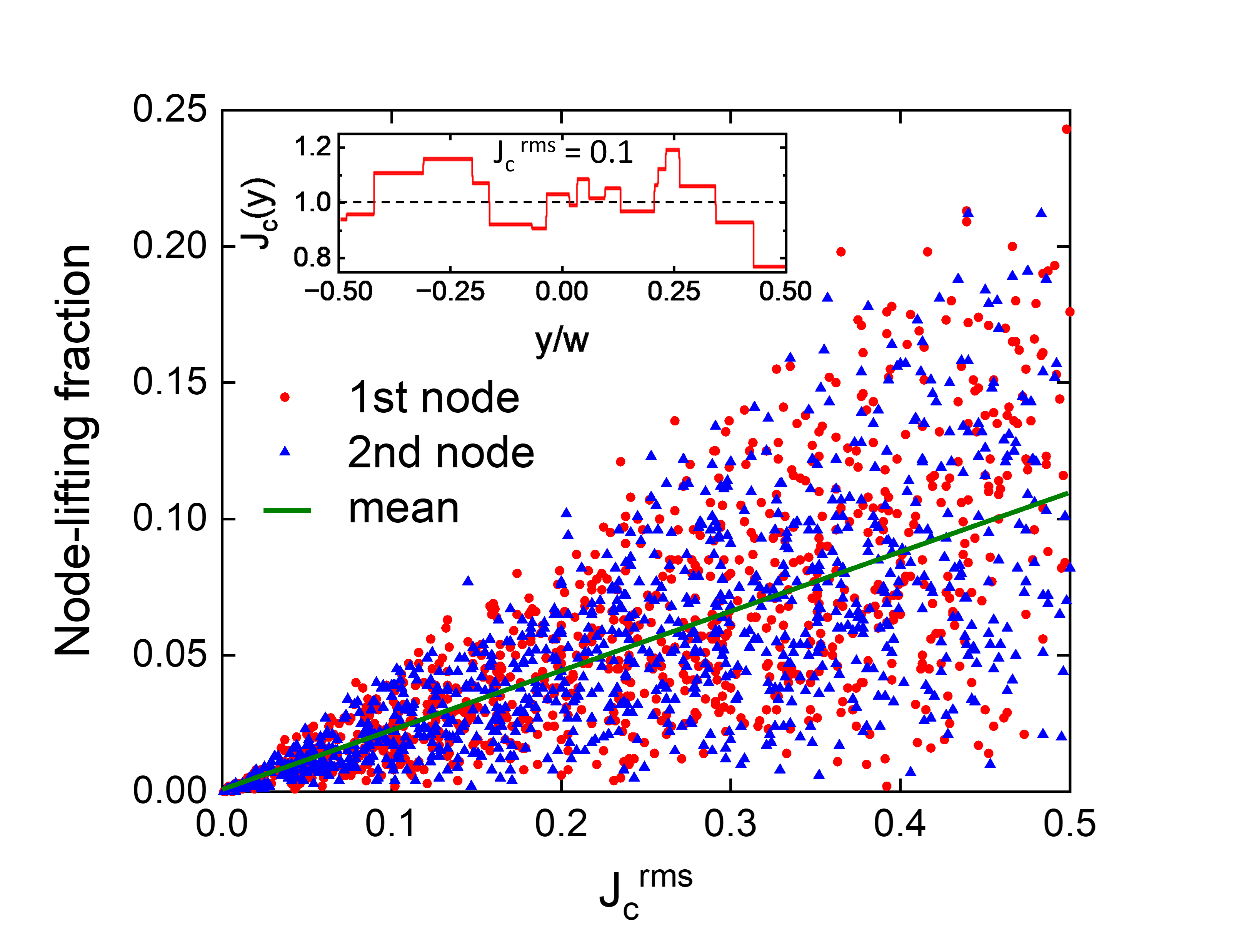}
    \caption{Simulations of the effects of random critical current disorder on the lifting of the first and second nodes in a sinusoidal Josephson junction, showing that both nodes lift on average by the same amount deviation.} 
    \label{fig:disorder1}
\end{figure}

\begin{figure}
    \centering
    \includegraphics[width=1.0\columnwidth]{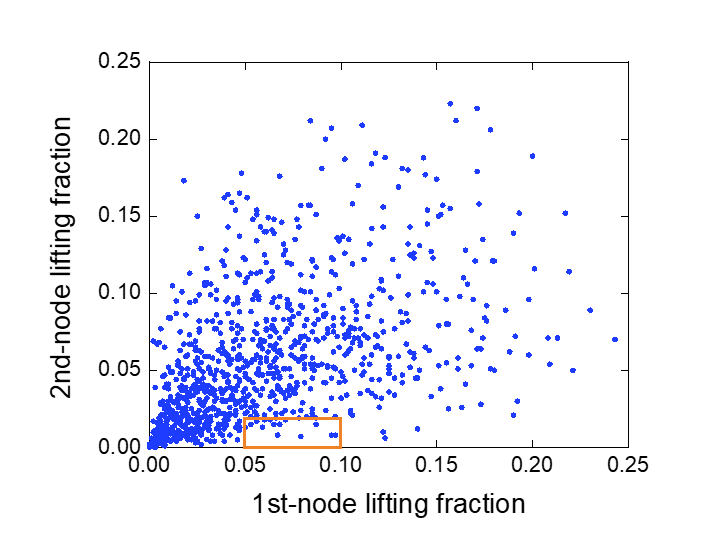}
    \caption{Simulations of the effects of random critical current disorder on the lifting of the first and second nodes in a sinusoidal Josephson junction, showing deviation from a Fraunhofer diffraction pattern; the small box shows regime of what we typically observe in the measurements on S-TI-S junctions.} 
    \label{fig:disorder2}
\end{figure}

To assess the impact of this level of critical current disorder, we have simulated the node-lifting in conventional Josephson junctions with a sinusoidal CPR as a function of junction critical current disorder. In the simulation of Fig. \ref{fig:disorder1}, we plot the lifting of the first and second node in the simulated diffraction pattern as a function of the RMS  critical current disorder for 1000 critical current distributions randomized in both the local width along the junctions and the magnitude. The range of the disorder is up to $50\%$ of the average critical current.  We see that the average node-lifting increases proportionately with the RMS strength of disorder, but for the typical geometry disorder $\sim5\%$ that we expect in our junction based on geometric effects, the junctions we measure should only have an average lifting of a few percent. Further, in Fig. \ref{fig:disorder2}, we plot the lifting of the first and second node for each of our simulations. This predicts that the first and second node lifting from random disorder should be comparable, and only  very few cases (those within the box indicated) should exhibit the significant first node lifting of $10\%$ and the zero or small second nodes that we frequently observe.  

\begin{figure}
    \centering
    \includegraphics[width=1.0\columnwidth]{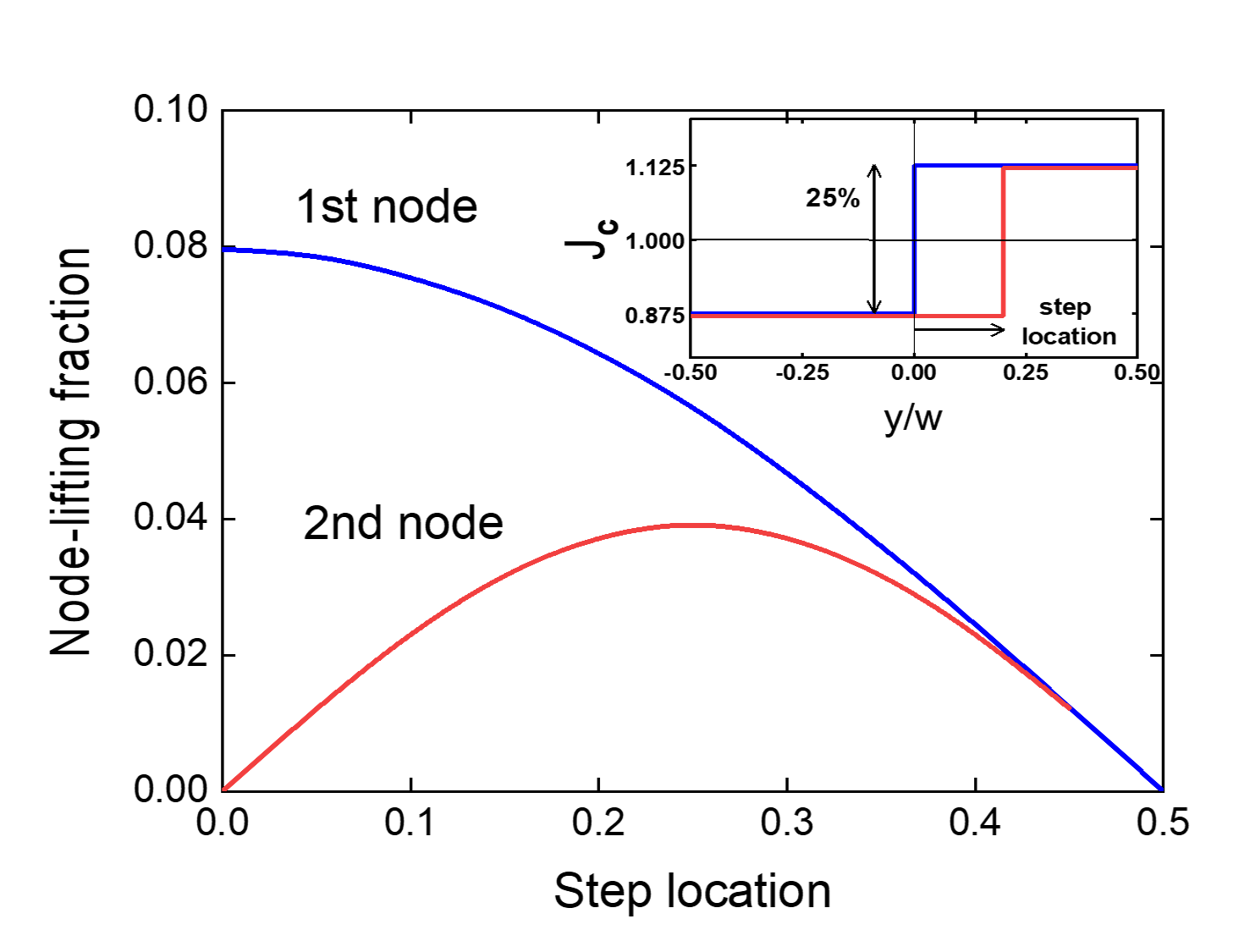}
    \caption{Simulations of the effects of an abrupt step in the junction critical current density on the lifting of the first and second nodes in a sinusoidal Josephson junctions.  This only agrees with our observations if the step is very close to the center of the junction.} 
    \label{fig:disorder3}
\end{figure}

\begin{figure}
    \centering
    \includegraphics[width=1.0\columnwidth]{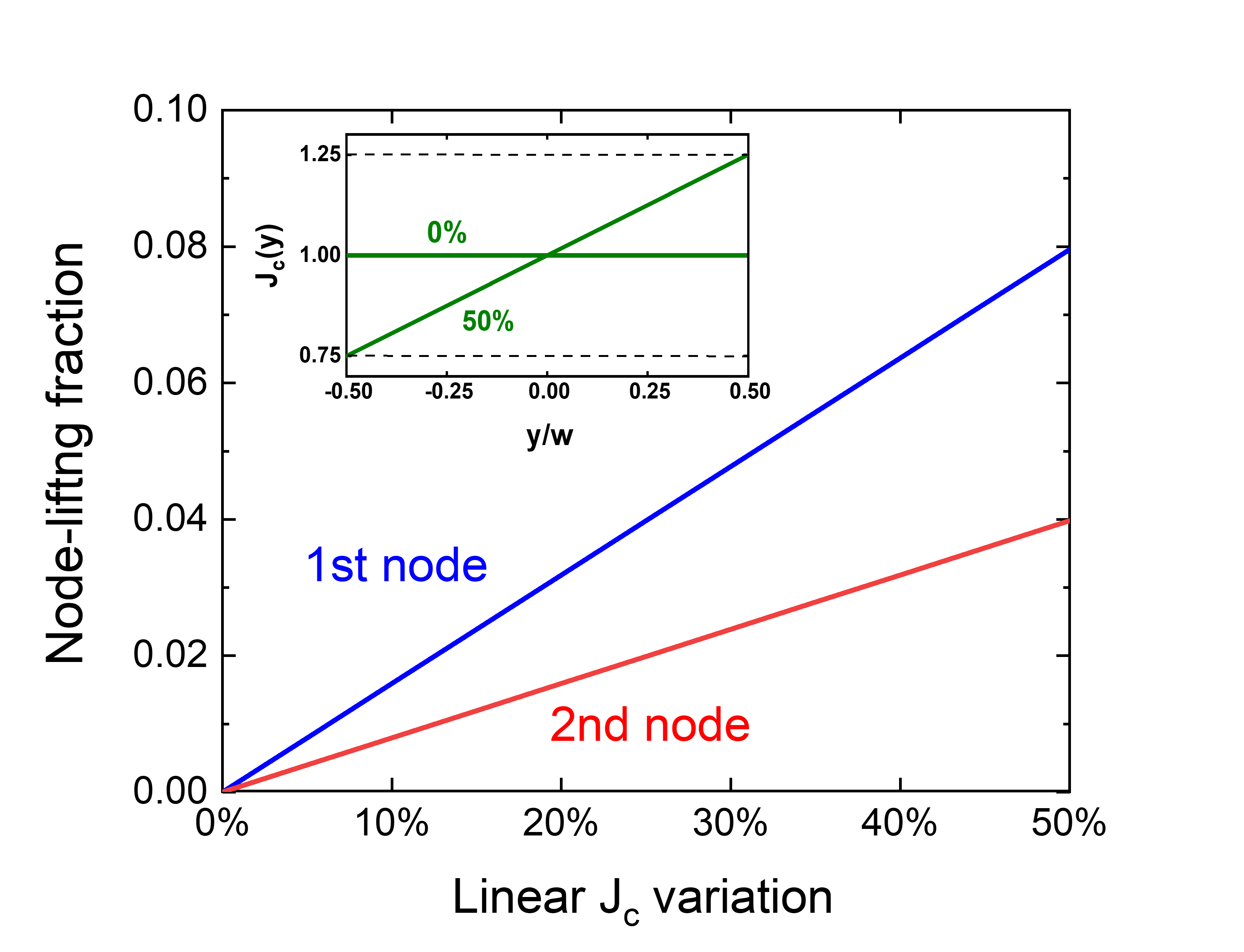}
    \caption{Simulations of the effects of uniform slope in the junction critical current density on the lifting of the first and second nodes in a sinusoidal Josephson junctions.} 
    \label{fig:disorder4}
\end{figure}

In addition to random disorder, we also consider several systematic variations in the critical current density along the width of the junction, $J_c(y)$. 

We then considered a step in $J_c$ at a specific location in the junction that might arise due to a fabrication glitch.  As shown in Fig. \ref{fig:disorder3}, such a step can lift the first node significantly while keeping the second node hard, but only if the step is very near the center of the junction. To get that result with the first node lifting comparable to what we see in our experiments, we put in a step of $50\%$ in $J_c$, a much larger deviation than we would expect.  

We also considered a linear variation in the current density across the width of the junction as might arise due to a widening of the barrier gap during  electron beam lithography.  To achieve a node-lifting of $8\%$, comparable to what we typically see in our devices, we needed to put in a variation of $25\%$ in the critical current, which for our spacing roughly corresponds to a comparable variation in the gap, far more than what is reasonable for our fabrication process. For this device, shown in Fig. \ref{fig:disorder4}, we would expect to see a lifting of the second node by $4\%$, which we do not observe.

Our conclusion from these simulations is that although junction critical current disorder can indeed affect the lifting of nodes in diffraction patterns in junction, we cannot possibly account for our body of data from this mechanism.

\subsection{Effects of parity fluctuations}

As mentioned above, one observation we deduce from our set of junction measurements is that although the first node is almost always lifted by a substantial amount, typically 5-15\%, the second node exhibits a broader distribution of values, i.e. there are a significant number of samples for which the critical current is not close to zero at the expected location of the second node.  Further, although it is challenging to measure the critical currents at higher node locations because of the small magnitudes, we do sometimes observe substantial lifting of these nodes, more so than we might expect from junction disorder. This prompted a study of the effects of parity fluctuations on the observed critical current.  

\begin{figure}[!ht]
    \centering   
    \includegraphics[width=1.0\columnwidth]{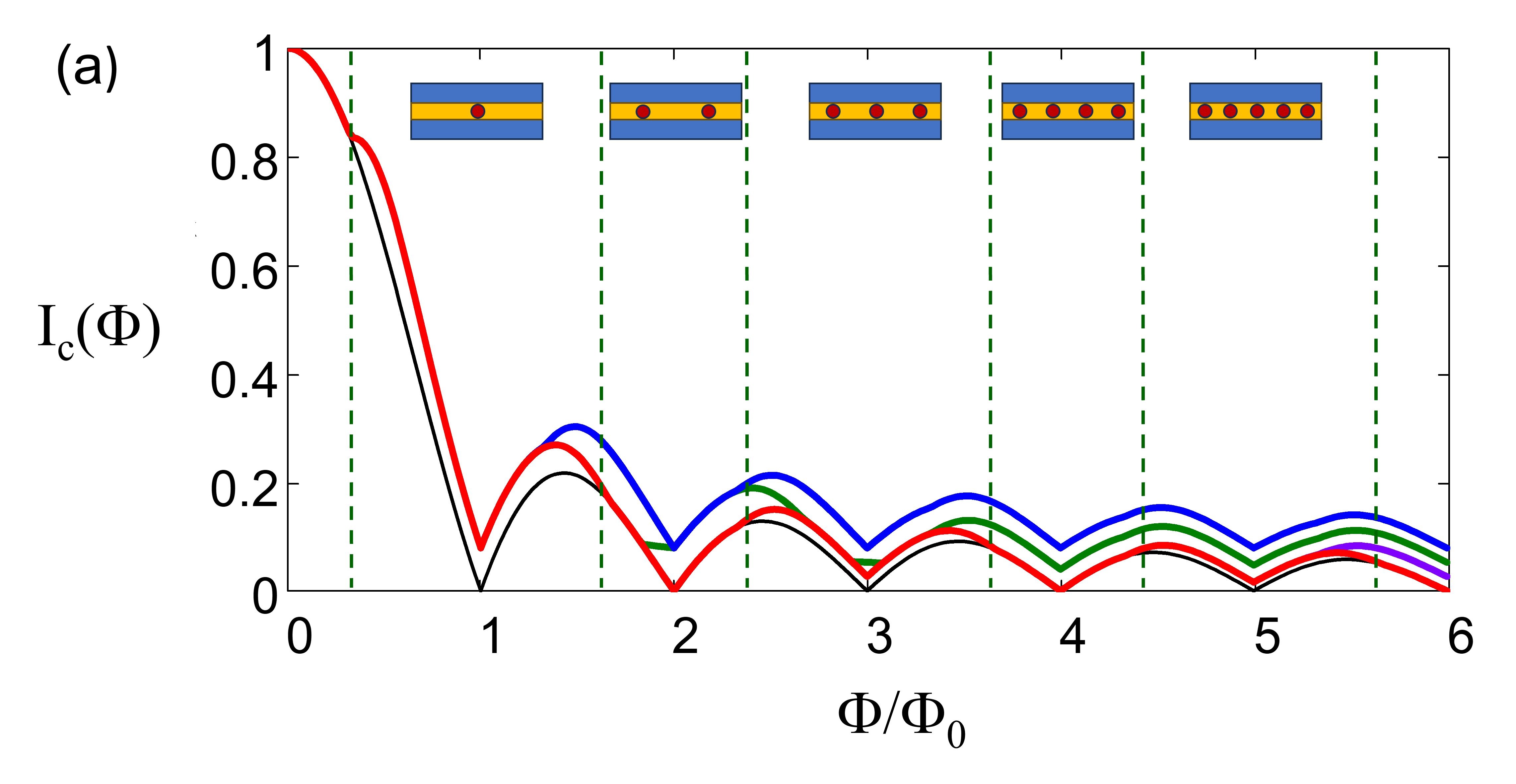}   
    \includegraphics[width=1.0\columnwidth]{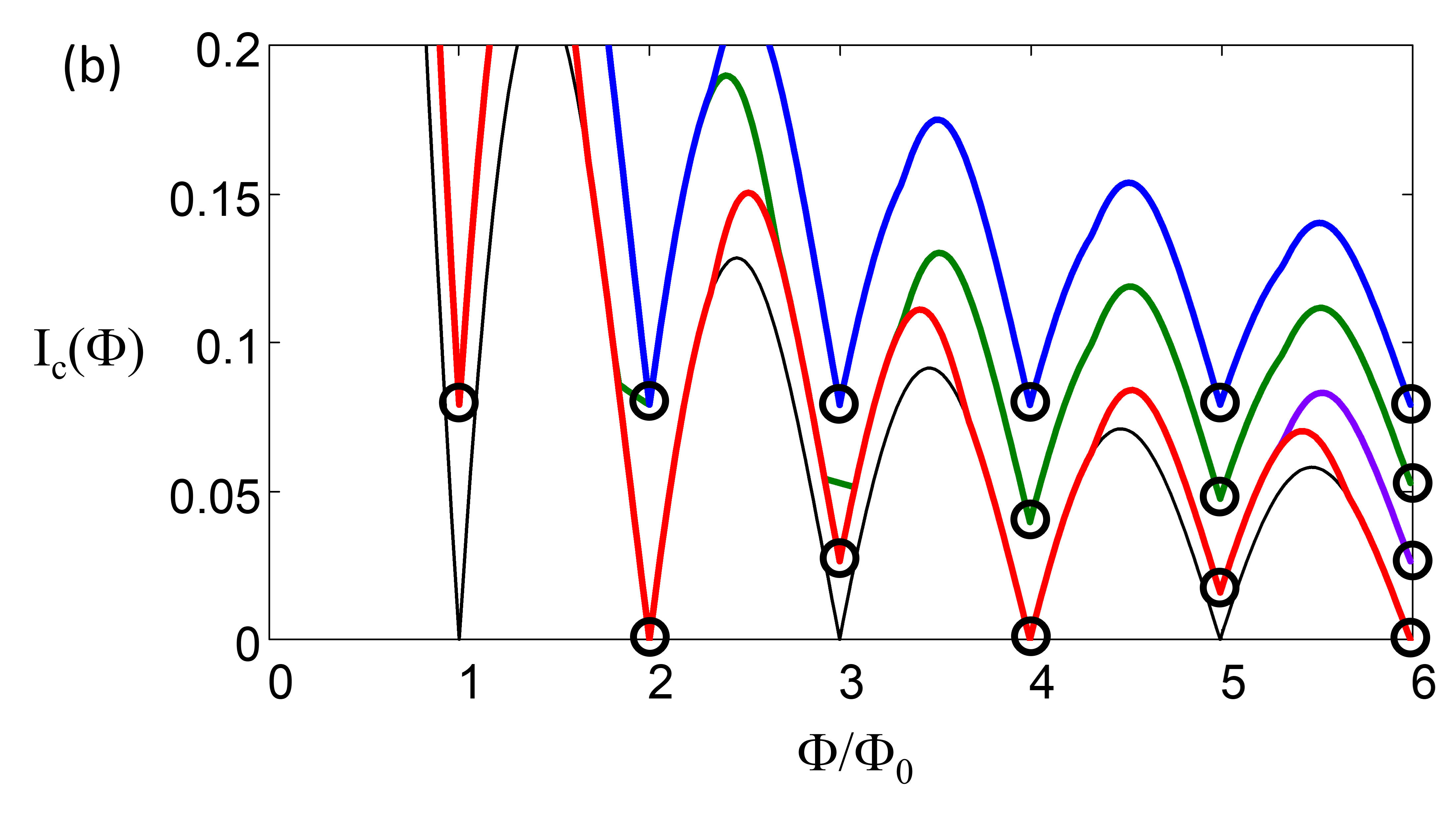}  
    \includegraphics[width=1.0\columnwidth]{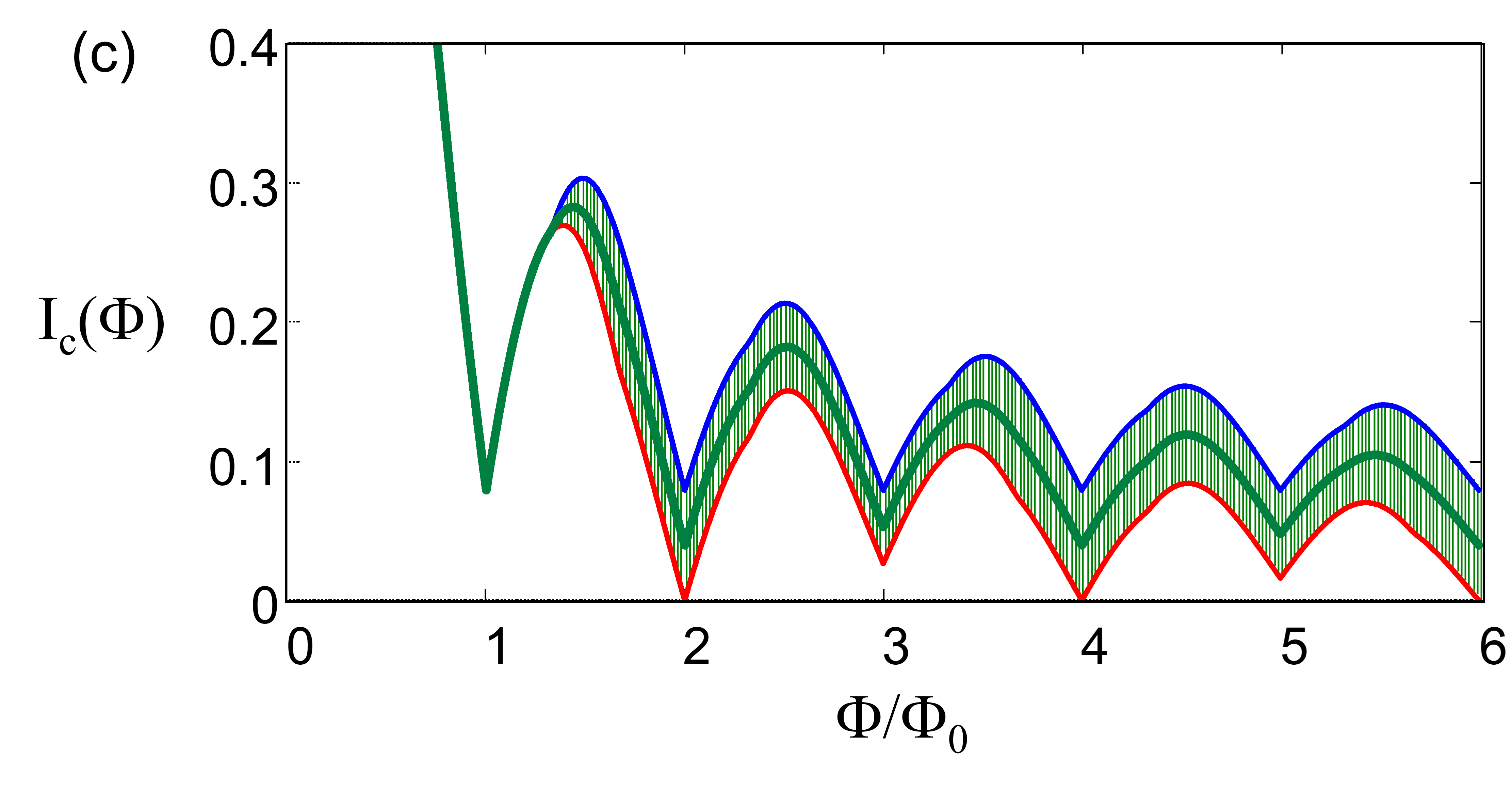} 
 
    \caption {(a) Diffraction patterns showing the effect of MBS parity changes.  BLACK: Fraunhofer diffraction pattern expected in the absence of Majoarana states.  RED:  Addition of discrete MBS.  GREEN:  Addition of a single parity change. BLUE: Addition of parity changes in alternating MBS.  The effect of parity fluctuations is to increase the critical current, with the measured diffraction pattern being a time-average of these curves. The pictures above show the number of MBS in the junction as a function of the applied magnetic field. (b) Expansion of the previous graph to show the allowable node heights in the presence of parity fluctuations.  (c) Averaged diffraction patterns in the presence of parity fluctuations. The green shaded areas shows the range of accessible curves for various parity transition rates.  The solid green curve is the diffraction patterns expected if the parity rate is comparable to the measurement rate set by the voltage threshold.} 
    \label{fig:Parity curves}
\end{figure}

Our picture is as follows.  We assume that although the MBS bound to Josephson vortices have a specific parity, there is a probability that the parity may change over time as result of interactions with quasiparticles, a phenomenon know as quasiparticle poisoning from which the MBS-pair parity is not protected. In the diffraction pattern measurement, which is deduced from a time-average of the differential voltage across the junction, we are sensitive to the critical current for different parities of the junction. To assess this, we calculate the diffraction pattern under the influence of non-uniform parity terms in Eqs. \ref{Eq:cpr} and \ref{Eq:alpha}. Fig. \ref{fig:Parity curves}(a) shows several calculated diffraction patterns when MBS inside the junction have different parities. At the top of this curve we indicate the number of Josephson vortices and MBS bound to them as a function of the applied flux. We see that the diffraction pattern does not change in shape up to and beyond the first node when there is zero or only one MBS in the junction.  In this regime, a parity transition does not change the critical current because the local phase can adjust to maintain the critical current at the same value.  However, for two or more MBS, we see that multiple values of critical current of the S-TI-S junction can exist at the same magnetic field and therefore parity transitions can modify the diffraction pattern.  Other nodes of the diffraction pattern can also be lifted up further in experiments in which the parity states of the MBS can vary across the junction or change with time.  

We expand this plot in Fig. \ref{fig:Parity curves}(b) to show more clearly the possible values of the critical current at the nodes.  We see that there are two possible node values at $\Phi$=2 or $\Phi$=3, three possible states at $\Phi$=4 or $\Phi$=5, and in general $integer(\Phi/2 +1)$ possible states at flux $\Phi$.  Since the diffraction pattern is determined as the time-average of the current at the threshold voltage of the junction, the measured diffraction pattern will be an average between the diffraction pattern for different parity states.  This is shown in Fig. \ref{fig:Parity curves}(c) in which the pattern will range in the green shaded regime bracketed by the curves for no parity transitions and fully randomized parity states. The solid green curve is an average between the two extremes expected if the parity transition rate is comparable to the Josephson frequency at the measurement threshold voltage. As noted above, this is typically in the 100 MHz to 1 GHz range. This diffraction pattern resembles some of the curves we measure showing an odd-even variation but in which all nodes are lifted, suggesting that parity fluctuations may explain the observed behavior.  

\section{Conclusions}
In this paper, we measured a collection of S-TI-S lateral Josephson junction critical current diffraction patterns and compared the results to a CPR model of this system that incorporates a uniform sinusoidal Cooper pair supercurrent and MBS localized on Josephson vortices that contribute a $4\pi$-periodic sin$(\phi/2)$ current. Taken as a whole, the data generally agrees with the key predictions of the model which include lifting of odd-numbered nodes and distinct features in the diffraction pattern at the applied magnetic fields at which Josephson vortices are expected to enter the junction. 

We emphasize, however, that individual samples can be affected by critical current disorder and MBS parity fluctuations, so  there is a wide range of observed diffraction patterns. To characterize these phenomena, we presented simulations designed to show the effect of these phenomena on the critical current. The key takeaway from the critical current disorder study is that although this effect can lift all nodes, it would take far more disorder than we have in our sample to generate a noticeable amount of node-lifting, and therefore this cannot be the sole origin of what we observe. On the other hand, we find that parity fluctuations are expected to create diffraction patterns that average over accessible parity distributions, the effect of which is to slightly lift all nodes, in particular the even nodes which would otherwise be at zero. The effectiveness of this lifting depends on the ratio of the characteristic measurement time, set by the Josephson frequency at the measurement threshold voltage, to the intrinsic parity lifetime. This is in agreement with observed diffraction patterns.   

None of this is by itself direct evidence for the existence of MBS.  These can ultimately only be verified by a demonstration of parity changes via MBS braiding. However, the general agreement of our measurements to the proposed model is intriguing and motivates us to further understand the physics of this system via direct current-phase relation measurements and by working toward achieving braiding by the exchange and subsequent readout of the parity state of MBS in this system, as we have discussed previously\cite{Hegde2019}. Such experiments are in progress.  It is also motivating work in our group to  explore the mechanism that affects the parity transition rates that will limit these measurements and the functionality of the S-TI-S junction platform. We are currently doing that via measurements of critical current distribution that should reveal the existence of parity fluctuations and provide a way to measure the parity rate, as we described in detail in previous work\cite{Abboud}.  

We continue to be intrigued and challenged by the possibility of observing the physics of MBS excitations in the S-TI-S system and exploiting it for the topologically-protected manipulation of quantum states.  

\section*{}

\section*{ACKNOWLEDGMENTS}
 We have benefited from many insightful discussions with Smitha Vishveshwara, Jim Eckstein, Alexey Bezryadin, Liang Fu, Jay Sau, and Jason Alicea.  
 
 We acknowledge the support of the National Science Foundation through grant DMR-2004825 and the Quantum Leap Challenge Institute for Hybrid Quantum Architectures and Networks grant OMA-2016136. The work at Rutgers University is supported by National Science Foundation’s DMR2004125, Army Research Office’s W911NF2010108, and MURI W911NF2020166, and the center for Quantum Materials Synthesis (cQMS), funded by the Gordon and Betty Moore Foundation’s EPiQS initiative through grant GBMF10104.

\bibliographystyle{apsrev}
\bibliography{STIS_prb_ref}

\end{document}